\documentclass[twocolumn]{aastex62}
\submitjournal{The Astrophysical Journal}
\usepackage{graphicx}
\usepackage{float}
\usepackage{amsmath}
\usepackage[caption=false]{subfig}
\usepackage[utf8]{inputenc}

\shorttitle{The Crab Nebula above 100 TeV}
\shortauthors{HAWC Collaboration}

\begin{document}
\title{\replaced{Measurement of the Crab Nebula at the Highest Energies with HAWC}{Measurement of the Crab Nebula Spectrum Past 100 TeV with HAWC}}
\author{A.U.~Abeysekara}
\affiliation{Department of Physics and Astronomy, University of Utah, Salt Lake City, UT, USA }
\author{A.~Albert}
\affiliation{Physics Division, Los Alamos National Laboratory, Los Alamos, NM, USA }
\author{R.~Alfaro}
\affiliation{Instituto de F\'{i}sica, Universidad Nacional Autónoma de México, Ciudad de Mexico, Mexico }
\author{C.~Alvarez}
\affiliation{Universidad Autónoma de Chiapas, Tuxtla Gutiérrez, Chiapas, México}
\author{J.D.~Álvarez}
\affiliation{Universidad Michoacana de San Nicolás de Hidalgo, Morelia, Mexico }
\author{J.R.~Angeles Camacho}
\affiliation{Instituto de F\'{i}sica, Universidad Nacional Autónoma de México, Ciudad de Mexico, Mexico }
\author{R.~Arceo}
\affiliation{Universidad Autónoma de Chiapas, Tuxtla Gutiérrez, Chiapas, México}
\author{J.C.~Arteaga-Velázquez}
\affiliation{Universidad Michoacana de San Nicolás de Hidalgo, Morelia, Mexico }
\author{K.P.~Arunbabu}
\affiliation{Instituto de Geof\'{i}sica, Universidad Nacional Autónoma de México, Ciudad de Mexico, Mexico }
\author{D.~Avila Rojas}
\affiliation{Instituto de F\'{i}sica, Universidad Nacional Autónoma de México, Ciudad de Mexico, Mexico }
% ...

\collaboration{HAWC Collaboration}
\correspondingauthor{Kelly Malone, Samuel Marinelli}
\email{kmalone@lanl.gov, marine20@msu.edu}
\begin{abstract}
We present TeV gamma-ray observations of the Crab Nebula, the standard reference source in ground-based gamma-ray astronomy, using data from the High Altitude Water Cherenkov (HAWC) Gamma-Ray Observatory. In this analysis we use two independent energy-estimation methods that utilize extensive air shower variables such as the core position, shower angle, and shower lateral energy distribution. In contrast, the previously published HAWC energy spectrum roughly estimated the shower energy with only the number of photomultipliers triggered. This new methodology yields a much improved energy resolution over the previous analysis and extends HAWC's ability to accurately measure gamma-ray energies well beyond 100 TeV. The energy spectrum of the Crab Nebula is well fit to a log parabola shape $\left(\frac{dN}{dE} = \phi_0 \left(E/\textrm{7 TeV}\right)^{-\alpha-\beta\ln\left(E/\textrm{7 TeV}\right)}\right)$ with emission up to at least 100 TeV. For the first estimator, a ground parameter that utilizes fits to the lateral distribution function to measure the charge density 40 meters from the shower axis, the best-fit values are $\phi_o$=(2.35$\pm$0.04$^{+0.20}_{-0.21}$)$\times$10$^{-13}$ (TeV cm$^2$ s)$^{-1}$, $\alpha$=2.79$\pm$0.02$^{+0.01}_{-0.03}$, and $\beta$=0.10$\pm$0.01$^{+0.01}_{-0.03}$. For the second estimator, a neural network which uses the charge distribution in annuli around the core and other variables, these values are $\phi_o$=(2.31$\pm$0.02$^{+0.32}_{-0.17}$)$\times$10$^{-13}$ (TeV cm$^2$ s)$^{-1}$, $\alpha$=2.73$\pm$0.02$^{+0.03}_{-0.02}$, and $\beta$=0.06$\pm$0.01$\pm$0.02. The first set of uncertainties are statistical; the second set are systematic. Both methods yield compatible results.  These measurements are the highest-energy observation of a gamma-ray source to date.
\end{abstract}

\section{Introduction}

The atmosphere is opaque to high-energy gamma rays; this means that they cannot be directly detected from the Earth's surface. Instead, these gamma rays interact with the atmosphere, initiating extensive air showers (EASs) that consist mainly of relativistic electrons, positrons, and photons.

The first gamma-ray/atmospheric interaction creates an electron-positron pair, which then creates additional gamma rays through the Bremsstrahlung process. This cycle repeats several times, with the total number of particles in the shower increasing exponentially. Due to conservation of energy, the average energy of each particle decreases as the number of particles increases. Eventually, the electron-positron pairs will reach the critical energy, where the radiative losses are equal to collisional energy losses and the shower begins to die out. This point is known as the ``shower maximum''. For a review on air shower development, see \cite{Matthews2005}.

Different types of ground-based gamma-ray detectors take different approaches in estimating the energy of the primary gamma ray of the EAS. The charged particles in the shower create Cherenkov light in the air as they travel to ground level. Imaging atmospheric Cherenkov telescopes (IACTs) work by detecting this Cherenkov light. Variables such as the image amplitude, the distance between the image and the center of the camera, the distance between the telescope and the shower axis, and the estimated height of the shower maximum are used to obtain gamma-ray energy estimates~\citep{Hofmann2000}. Techniques used may include look-up tables~\citep{Holder2015} or template-based analyses~\citep{Bohec1998,Parsons2014}. 

EAS arrays work by detecting the shower particles that reach ground level. Energy must be reconstructed using only this information. Because of this, it is a challenge to measure gamma-ray energies using an EAS array. For $\sim$1 TeV showers, the shower maximum occurs, on average, at a higher altitude (several tens of km) than ground level.  Shower fluctuations mostly stemming from the depth of the first interaction in the atmosphere limit the energy resolution.  As the energy of the primary gamma ray increases, shower maximum becomes closer to ground level. This leads to better energy resolution.

The simplest way to obtain a gamma-ray energy estimate with an EAS array is to count the number of detector elements triggered during an air shower event. This method was used by the Milagro experiment~\citep{Abdo2012}, among others. However, this parameter is typically only weakly correlated with energy as it does not take into account some important variables: the zenith angle of the event, the distance from the air shower core to the array, and how well-contained the shower is within the array. The zenith angle determines how much atmosphere a shower travels through on its way to the Earth's surface, while the distance to the air shower core determines the overall level of signal detected. The containment of the shower within the array can lead to a lack of dynamic range at the highest energies. Above some energy threshold, every detector element may be triggered. At this point, it becomes impossible to estimate the gamma-ray energy just from the percentage of detector elements hit.

Some EAS arrays have utilized the normalization of the lateral distribution function (LDF) of the shower, as this quantity compensates for fluctuations in the lateral distribution.  The LDF method is inspired by a technique originally proposed in the 1970s \citep{Hillas1971} and used by large cosmic-ray experiments such as the Pierre Auger Observatory and KASCADE-Grande \citep{Newton2007, Apel2016}, but with some modifications made for the typically smaller physical size of gamma-ray EAS arrays compared to arrays designed to detect cosmic rays. One implementation of this method is used by Tibet~\citep{Kawata2017}, which uses the particle density 50 meters from the air shower axis to obtain an estimate of the gamma-ray energy.  

In this paper, we describe two new gamma-ray energy estimation algorithms developed for the High Altitude Water Cherenkov (HAWC) Gamma-Ray Observatory. These methods extend the dynamic range of the experiment above 100 TeV, which is important for analyses such as PeVatron searches and studies of Lorentz-invariance violation.

The two methods are validated on the Crab pulsar wind nebula. In 1989, this source became the first to be convincingly detected in TeV gamma rays~\citep{Weekes1989}. Since then, it has been observed by nearly all TeV gamma-ray experiments. As the steady source with the brightest integral flux above 1 TeV, it is often used as a reference source.  



Even though the Crab Nebula has been extensively studied, observations at the highest energies ($>$ 50 TeV) are sparse. This is due to the source's small flux in this energy range. Two interesting results in the literature are the HEGRA detection \citep{Aharonian2004}, which includes a 2.7$\sigma$ detection above 56 TeV, and the limits set by the Tibet Air Shower Array above 100 TeV \citep{Amenomori2015}.  The Crab spectrum presented here extends roughly \replaced{twice}{three times} as high in energy as HAWC's previously published Crab spectrum~\citep{Abeysekara2017}.

The paper organization is as follows: section \ref{sec:hawc} provides a basic description of HAWC. Section \ref{sec:est} describes the two independent gamma-ray energy estimation algorithms. Section \ref{sec:crab} shows the application of these energy estimation algorithms to the Crab Nebula, while Section \ref{sec:disc} discusses possible implications of these results.

\section{The HAWC observatory}\label{sec:hawc}
HAWC is a gamma-ray detector located in the state of Puebla, Mexico, at an elevation of 4100 meters.  It consists of 300 water Cherenkov detectors, each outfitted with four PMTs (three of which are 8-inch and one of which is a 10-inch higher-quantum efficiency PMT). When the gamma rays in the EAS hit the water, they produce electron-positron pairs. All of the charged particles from the air shower then produce Cherenkov radiation which detected by the PMTs. HAWC's design, data acquisition architecture, and reconstruction methods are described extensively in ~\cite{Smith2015, Abeysekara2017, Abeysekara2018}.  HAWC is optimized to detect gamma rays in the 100 GeV to 100 TeV range, although it can detect emission above 100 TeV.  HAWC is located at 19$^\circ$ North, which is nearly the optimal viewing angle for observations of the Crab Nebula.  

\begin{figure*}
\includegraphics[width=0.5\textwidth]{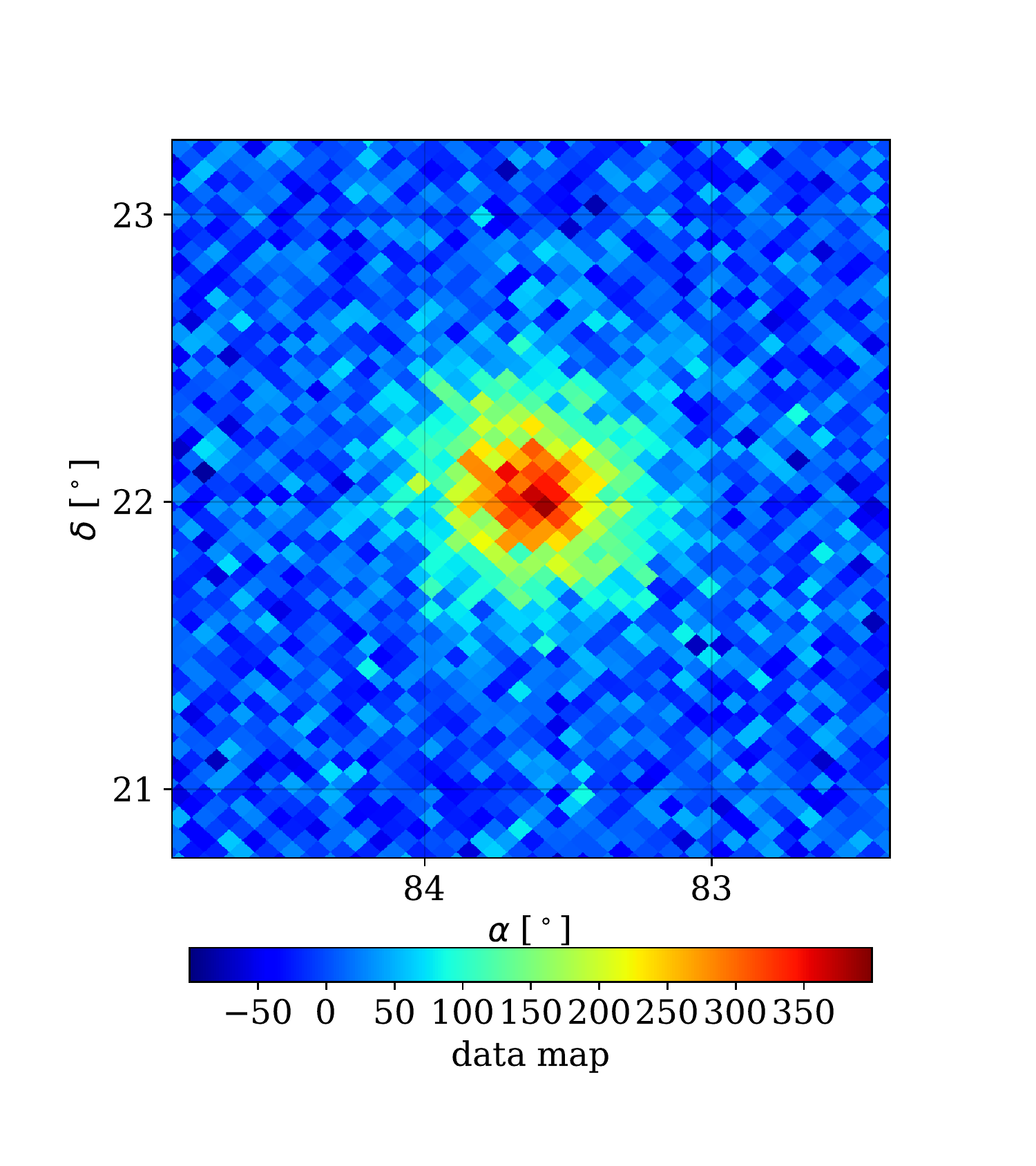}
\subfloat{\includegraphics[width=0.5\textwidth]{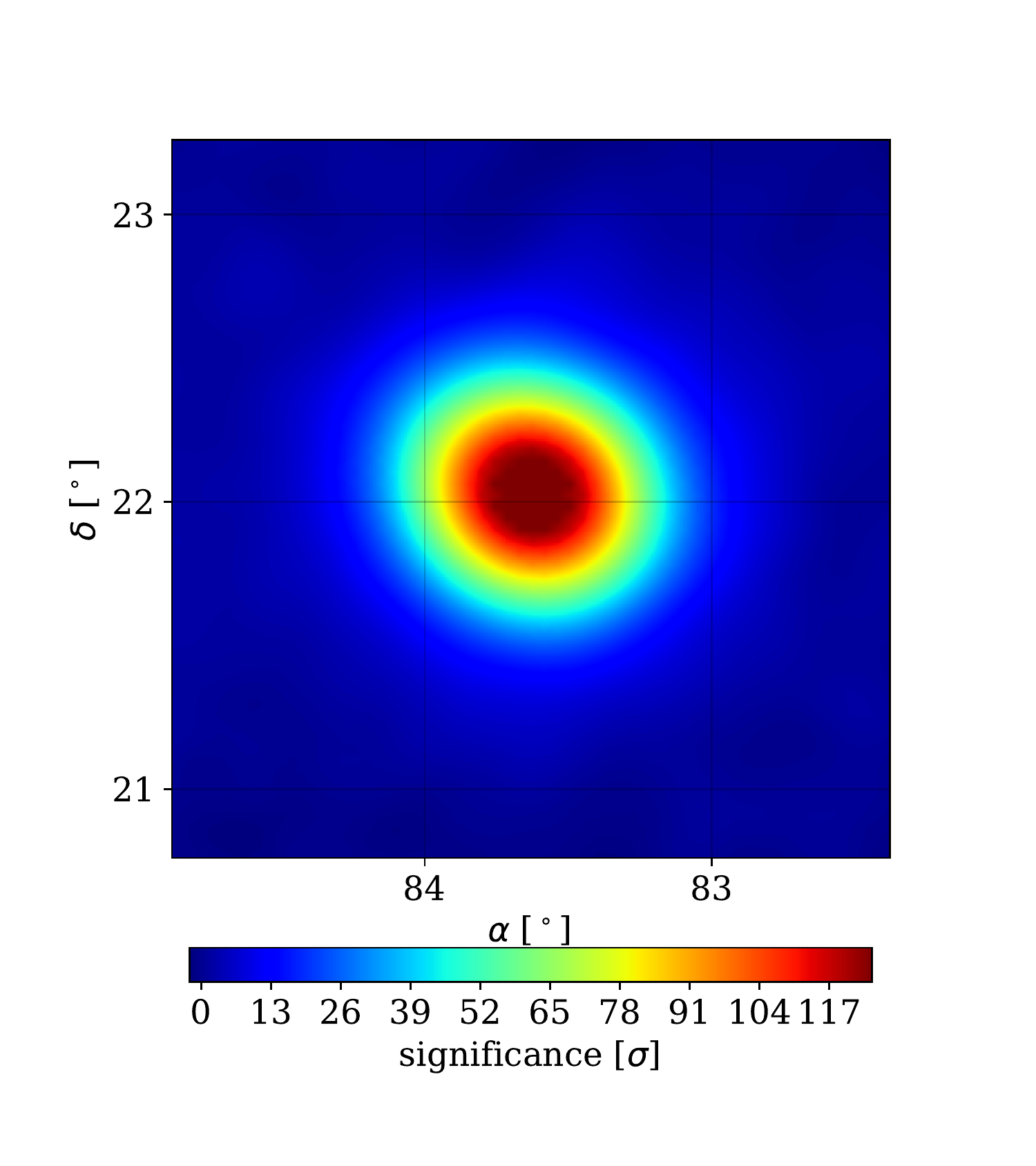}}
\caption{Left: A map of the Crab Nebula region depicting excess counts above 1 TeV in reconstructed energy. 837.2 days of data are included. Gamma/hadron separation cuts and background subtraction have already been applied.  Right: The corresponding significance map. Significance is calculated using a likelihood framework. The maximum significance is 139$\sigma$. The significance map has been smoothed for presentation purposes. } 
\label{fig:countsmap}
\end{figure*}

\added{HAWC records air shower events at a rate of 25kHz. Most of the triggers are due to cosmic ray-induced air showers. Over the course of one day, only a few hundred of these air showers are gamma-like and coincident with the Crab Nebula. The exact number depends on the quality cuts chosen.  Figure \ref{fig:countsmap} shows a typical background-subtracted counts map of gamma-like events above 1 TeV in reconstructed energy from the direction of the Crab. The corresponding significance map is also shown, produced using the hypothetical point-source prescription described in \cite{Abeysekara2017a}. }

\section{Energy estimation}\label{sec:est}
HAWC's first published observations of the Crab Nebula \citep{Abeysekara2017}, as well as all of HAWC's subsequent gamma-ray focused publications up to this point, used an extremely simplistic energy estimator: the size of an air shower event was used as a proxy for energy. Events were placed in analysis bins (indexed here by $\mathcal{B}$) depending on what fraction of the PMTs in the array were triggered during the event. $\mathcal{B}$ is only weakly correlated with energy, as discussed in the introduction. In the last analysis bin, every PMT was triggered and the $\mathcal{B}$-based energy proxy saturated. This bin included nearly every event above $\sim$30 TeV, making it impossible to distinguish 30 TeV events from 100 TeV events. The Crab Nebula spectrum presented in \cite{Abeysekara2017} was only reported up to 37 TeV due to this saturation of the analysis at high energies. A complete energy analysis is presented here for the first time, allowing for the extension of the spectrum to much higher energies. This analysis uses two independent energy-estimation algorithms. This allows for cross-checking of results.  This is particularly important for the highest energies (i.e. above 100 TeV), which are inaccessible to \added{most} other currently operating gamma-ray detectors.

The two methods, the ground parameter (GP) and the neural network (NN), are described below. Throughout this section, $\hat{E}$ will refer to estimated energy while $E$ will refer to the \replaced{true energy from simulation}{simulated energy}. These two energy estimation methods were developed using HAWC's standard Monte Carlo simulation, which relies on Corsika v7.4000 \citep{Heck1998} to simulate the air showers and GEANT4 v4.10.00 \citep{Agostinelli2003} to propagate the secondary particles from those air showers through the HAWC detector to the photomultiplier tubes.  The data acquisition system is modeled by a custom piece of software called DAQSim.  
\subsection{Ground parameter algorithm}
\label{sec:gp}
The GP algorithm is based primarily on the charge density at a fixed optimal distance from the shower axis.  As discussed in the introduction, this is a robust estimator of the energy reaching the ground.  

The radius at which the uncertainty in the shower energy density is minimized (hereafter known as the ``optimal radius'') must be far from the air shower axis due to the presence of large shower-to-shower fluctuations that make energy estimation difficult, but it also must be close enough to the shower axis that the measured PMT signal is large enough that threshold effects in the electronics are not a concern.

To determine this optimal radius, the LDF is fit to a modified version of the NKG function. The canonical NKG function measures particle density. The signals in water Cherenkov detectors scale with energy deposited in the water. Therefore, a factor of $1/r$ is introduced to measure energy density rather than particle density. The lateral distribution of the energy is steeper by a factor of $1/r$ because the highest energy particles are less likely to be scattered away from the air shower axis~\citep{Kamata1958}. Note that this technique is similar to the method used by the Tibet air shower array~\citep{Kawata2017}. The main difference is that the Tibet implementation measures particle density, while this method measures energy density.

\begin{figure}
\centering
\includegraphics[width=0.48\textwidth]{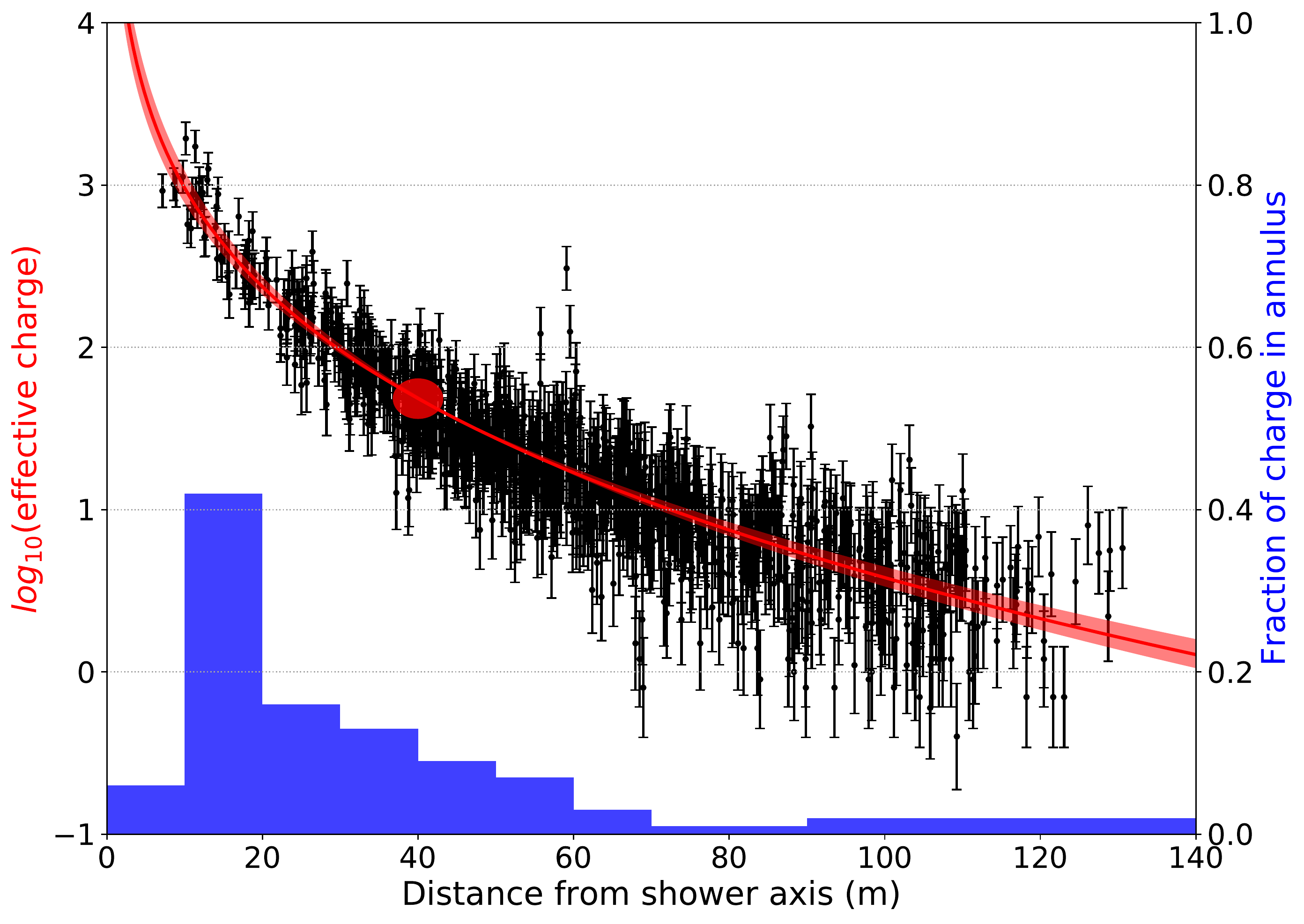}
\caption{A depiction of the information used by the two energy estimators. The black points show, for a single event, the log of the effective charge measured by each PMT as a function of the distance to the air shower axis. Effective charge introduces a scaling factor for the high quantum efficiency PMTs to place them on par with the other PMTs. The red line is the best fit to the NKG-like function, while the band was used to determine the optimal radius (see section \ref{sec:gp}).  The circle at $r = 40$ meters visually denotes the location the charge is measured at in the GP method, which is used along with the zenith angle to calculate the estimated energy. The blue histogram is the fraction of charge in several radial rings, which are used as some of the inputs to the NN. }
\label{fig:ldf}
\end{figure}

The LDF fit function, which gives the PMT signal (charge, hereafter called $sig_{r}$) as a function of several variables, is:

\begin{equation}
\label{eq:nkg}
\begin{split}
\log_{10}\!\left(sig_{r}\right) = A + s\left[\log_{10}\frac{r}{r_m} + \log_{10}\!\left(1 + \frac{r}{r_m}\right)\right]  \\
{}-3\log_{10}\frac{r}{r_m} - 4.5\log_{10}\!\left(1 + \frac{r}{r_m}\right).
\end{split}
\end{equation}
Here, $A$ is the logarithm of the amplitude of the fit, $s$ is related to the shower age, and $r_m$ is the Molière radius, which is $\sim$124 m at the HAWC site. \added{Since this is the NKG/r function instead of the NKG function, the amplitude of the fit absorbs some of the dependence on the shower age and a conversion is needed to calculate the true shower age.} $r$ is the distance from the PMT to the air shower axis. $A$ and $s$ are the only two free parameters in the fit. See Figure \ref{fig:ldf} for a depiction of a lateral distribution function fit to this NKG/$r$ function. When doing this fit, PMTs with zero signal are ignored.  

After obtaining the best fit, $s$ is varied by $\pm$10\%, and additional fits are performed, leaving the normalization free. This creates a band of fits (see Figure \ref{fig:ldf}). The location where the width of the band is the smallest is the point where the uncertainty in the signal is minimized. This distance is known as the optimal radius. For HAWC, this value is found to be $\sim$40 meters from the shower axis irrespective of zenith angle or primary-particle energy.  This mirrors the findings in \cite{Newton2007}, which notes that the optimal radius is almost solely a function of array geometry.

Equation \ref{eq:nkg} is evaluated at $r$=40 meters. This value is known as $\log_{10} sig_{40}$; it is then translated to energy. For a fixed primary energy, the signal measured on the ground varies strongly with zenith angle due to the differing amount of atmosphere that air showers entering at different angles must travel through. The formula must be parameterized by zenith angle: $\log_{10} \hat{E} = f(sig_{40},\theta)$.  The exact functional form of the fit is chosen empirically to provide a good match to simulated events:

\begin{equation}
\log_{10} \hat{E} = m\!\left(\theta\right) \log_{10} sig_{40} + c\!\left(\theta\right).
\end{equation}
Here, $m(\theta)$ is chosen to be a piecewise linear function and $c(\theta)$ is chosen to be a piecewise quadratic function.

It is important to note that 40 meters is only the mean optimal radius, and that for a given event the true optimal radius may be higher or lower. To quantify the effect of using one optimal radius for all events, the procedure above was repeated with the optimal radii set to both 30 and 50 meters. No systematic shifts in the assigned energy were observed. 

For the performance of the GP on simulation, see Section \ref{sec:performance}.

\subsection{Neural network algorithm}

    The NN energy-reconstruction algorithm employs an artificial neural network
    to estimate primary energies of photon events based on several quantities
    that are computed as part of HAWC's event reconstruction. The Toolkit for
    Multivariate Analysis (TMVA) NN implementation, described in
    \cite{Hocker:2007ht}, is used.

    The NN energy estimator uses a multilayer-perceptron architecture with two
    hidden layers and a logistic activation function. The first and second
    hidden layers have
    15 and 14 nodes respectively.
 
    The values of the 479 NN weights are chosen to minimize the error function
    
    \begin{equation}
        D\!\left(\mathbf{w}\right) \equiv \frac{1}{2} \sum_{i=1}^n u_i \left[
            \log_{10} \hat{E}\!\left(
                \mathbf{x}_i; \mathbf{w}
            \right) - \log_{10} E_i
        \right]^2.
    \end{equation}
    This is evaluated using Monte Carlo events,
    where $\mathbf{w}$ is the vector of NN weights, $n$ is the number of
    events,
    $u_i$ is the relative importance of the $i$th event, $\mathbf{x}_i$ is the
    vector of input variables for the $i$th event, $\hat{E}$ is the function
    returning an energy estimate for a given vector of inputs and vector of
    weights, and $E_i$ is the \replaced{Monte Carlo true energy}{simulated energy} of the $i$th event. The values of
    $u_i$ are chosen to resemble an $E^{-2}$ power
    law, which was selected because a NN trained on such a spectrum was found
    to produce a relatively constant RMS error
    between 1 and 100 TeV, as shown in Figure \ref{fig:rmserror}.
    The
    minimization of the error function is performed via the
    Broyden-Fletcher-Goldfarb-Shanno algorithm, described in
    \cite{Hocker:2007ht}.

    For the performance of the neural network on simulation, see Section
    \ref{sec:performance}.

    \subsubsection{Input variables}

        The NN input variables are chosen to describe three broad
        characteristics of the air shower: the amount of energy deposited in
        the
        detector, the extent to which the shower's footprint on the ground is
        contained within the detector, and the degree of attenuation of the
        shower
        by the atmosphere. The resulting algorithm can be thought of as a
        calorimetric measurement combined with corrections for the fraction of
        the shower not hitting the detector and for the atmospheric
        attenuation.

        Three quantities are used to infer the amount of energy deposited in
        the detector: the fraction of PMTs hit within the event, the fraction
        of tanks hit, and the logarithm of the normalization from the
        fit of the lateral distribution function. The lateral distribution function used is not the modified NKG described in Equation \ref{eq:nkg}; instead the Super Fast Core Fit (SFCF), described in \cite{Abeysekara2017}, is used. The SFCF uses a smoothed approximation to the NKG of Equation \ref{eq:nkg}.  All of the above parameters are positively
        correlated with the shower's primary energy.

        The fraction of the shower landing within the detector on the ground is
        inferred using the distance between the reconstructed core location of
        the shower and the center of the HAWC array.

        The atmospheric attenuation of the shower is quantified in two ways:
        using the cosine of the reconstructed zenith angle of the shower and
        using the shower's lateral charge distribution, which contains information
        about the
        shower age. The lateral distribution is passed to the NN in the form of
        ten input variables. The first nine of these variables consist of the
        fraction of the charge deposited in all PMTs during the event that
        lands within each of nine concentric annuli about the reconstructed
        shower axis. Each annulus has a width of 10 m. The last of
        these ten input variables is the fraction of the event's charge landing
        more than 90 m from the shower axis (see Figure \ref{fig:ldf}).

\subsection{Performance of the estimators}
\label{sec:performance} \label{sec:biasres}
\begin{figure*}
\includegraphics[width=0.5\textwidth]{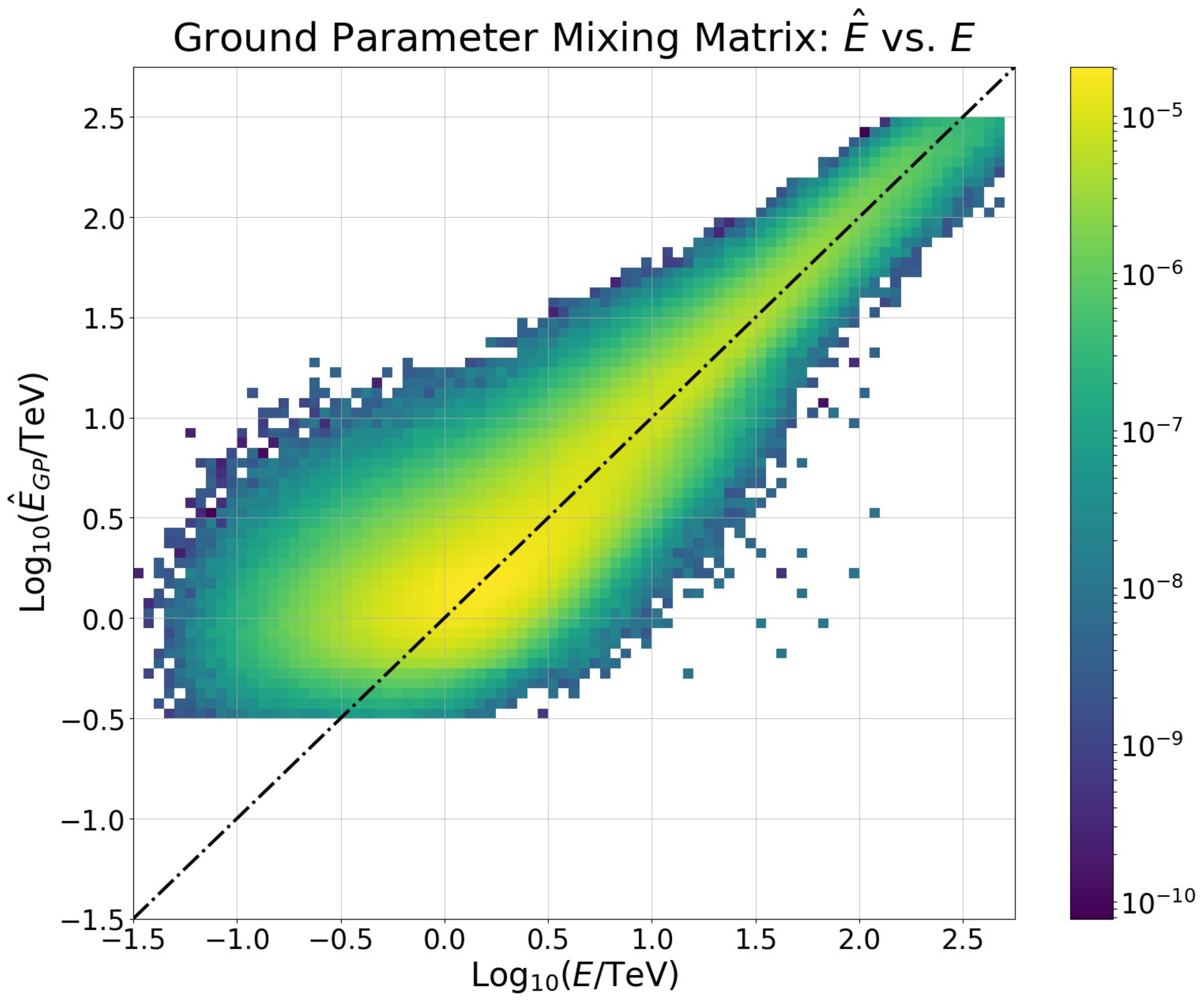}
\subfloat{\includegraphics[width=0.5\textwidth]{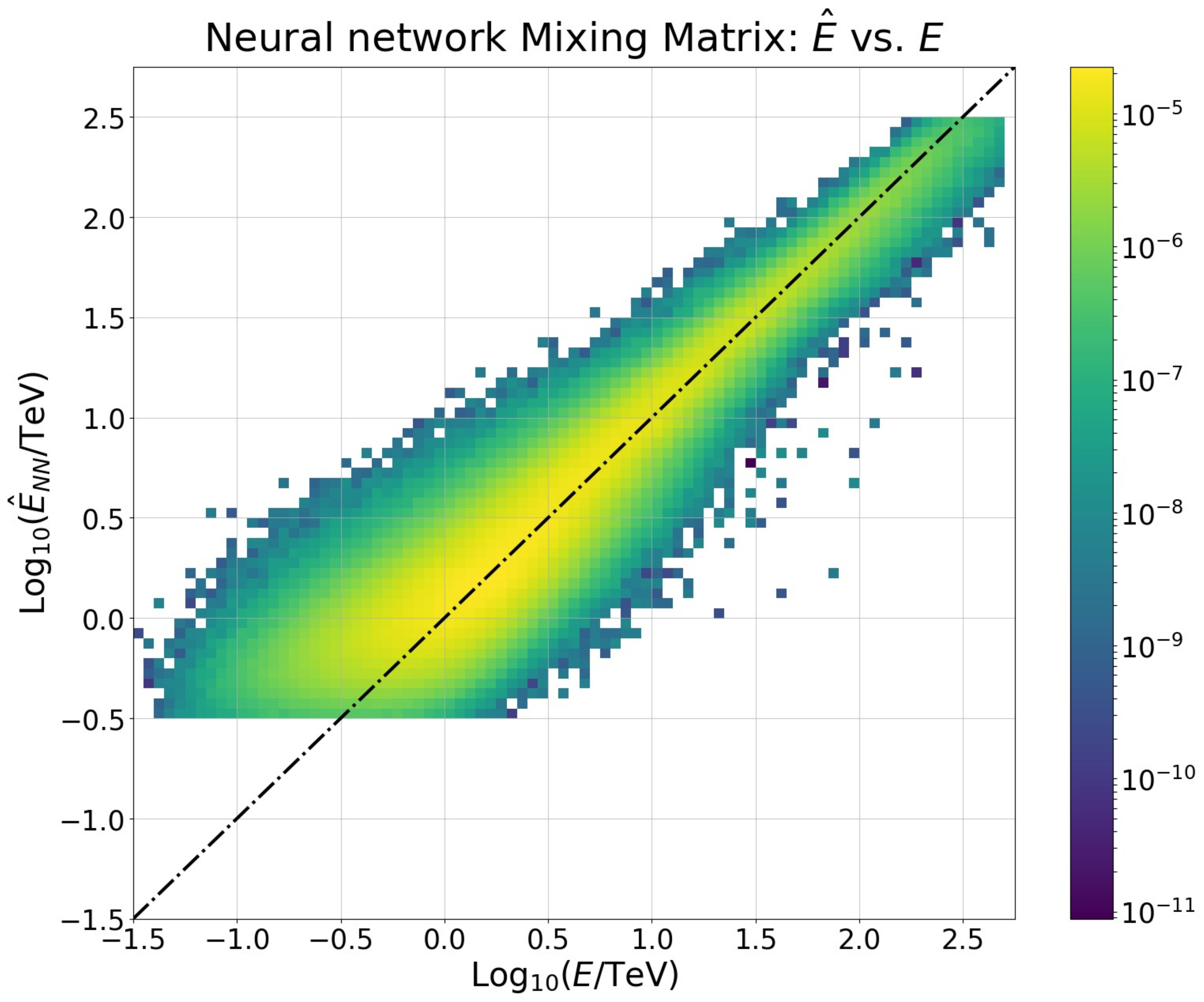}}
\caption{The mixing matrices for the GP (left) and NN (right) energy estimators. The dotted line is the identity line; events that fall along this line are reconstructed perfectly. Gamma/hadron separation cuts have been applied. The first energy bin starts at $log_{10}(\hat{E}/ \text{TeV}) =$ -0.5 and the last energy bin ends at $log_{10}(\hat{E}/ \text{TeV}) =$ 2.5, which accounts for the sharp features in the figures. }
\label{fig:performance}
\end{figure*}
The mixing matrices, which compare the energy estimate to the \replaced{true energy}{simulated energy}, can be seen in Figure \ref{fig:performance}. Each plot is normalized as a joint distribution in true and reconstructed energy, so that its two-dimensional integral is 1. This figure assumes an isotropic $E^{-2}$ spectrum of gamma rays; this assures that there are sufficient events at high energy to evaluate the performance. Several data quality cuts have been applied here: only simulated gamma-ray events whose shower core is successfully reconstructed on the HAWC array, have PMT signal in more than 6.7$\%$ of the active detectors in the array (corresponding to $\mathcal{B}$ bins 1 and above), and have a zenith angle of $< 45^\circ$ are used. Additionally, the events are selected to have a reconstructed zenith angle less than 0.75$^\circ$ from the true Monte Carlo value. To more accurately show what this \replaced{plot}{figure} would look like for data, gamma/hadron separation cuts have been applied to the simulated gamma-ray events.

\begin{figure}
\includegraphics[width=0.48\textwidth]{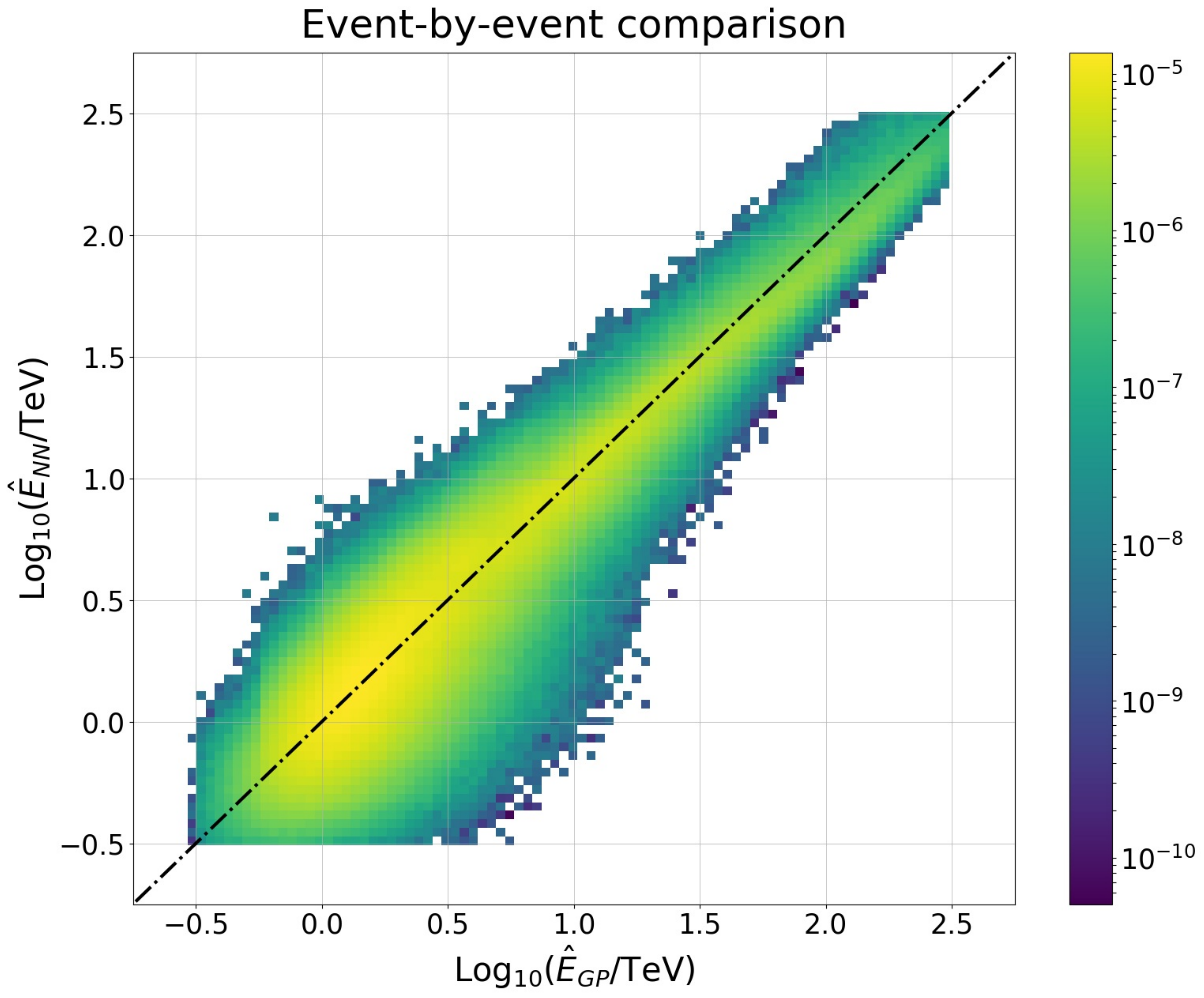}
\caption{Event-by-event comparison of the two estimators, for gamma-ray events that pass data quality cuts. The optimal gamma/hadron separation cuts differ for the two estimators; only events passing both sets of cuts are shown here.  The dotted line is the identity line. Events falling on this line have the same energy estimate regardless of which method is used.}
\label{fig:comparison}
\end{figure}

Figure \ref{fig:comparison} shows an event-by-event comparison of the two estimators. All of the quality cuts described in the preceding paragraph are also used here. The optimal gamma/hadron separation cuts are different for each estimator. Only events passing both sets of gamma/hadron cuts are shown.

\begin{figure}
\centering
\includegraphics[width=0.48\textwidth]{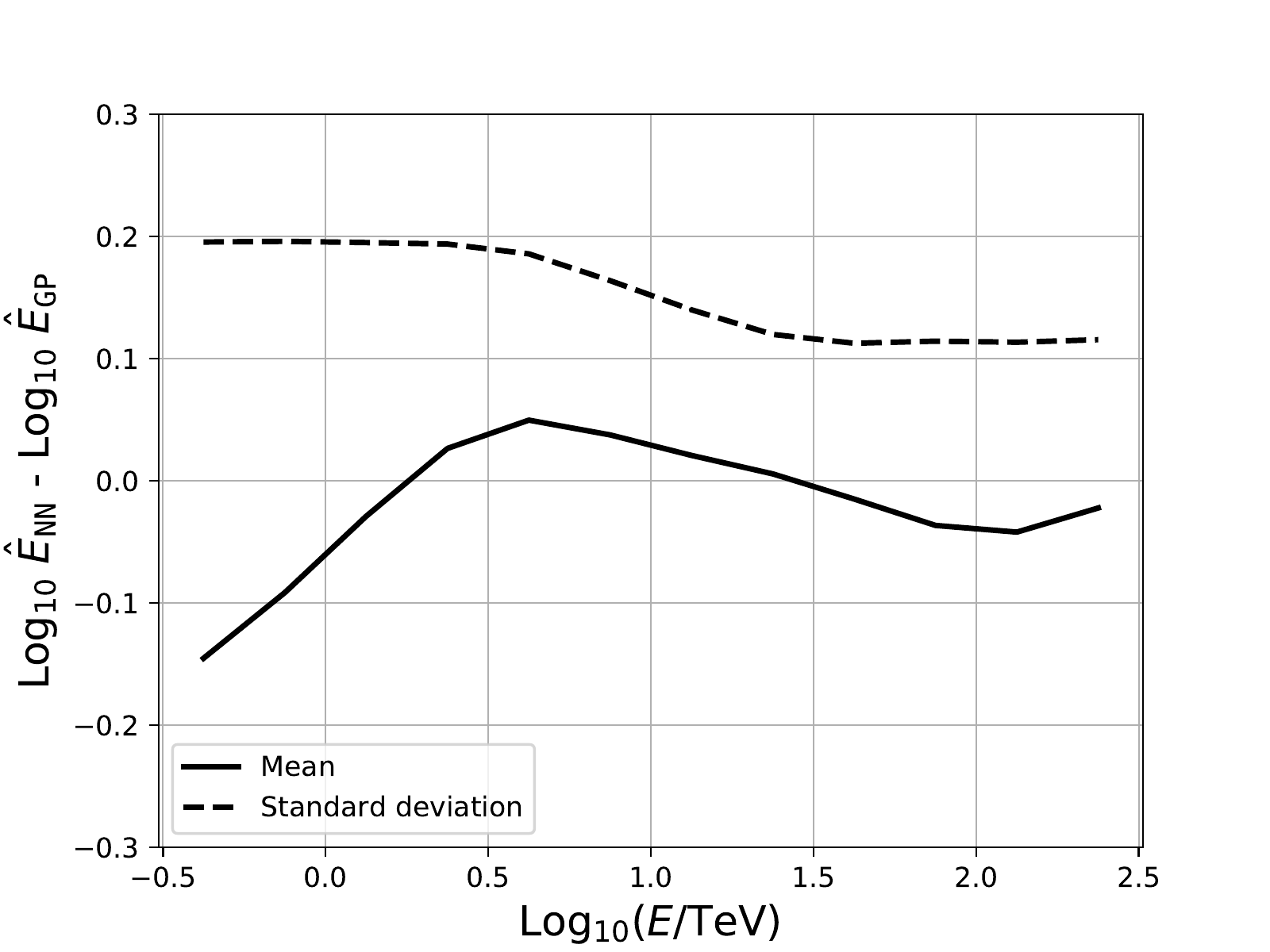}
\caption{The difference between the energy estimates as a function of $E$. Gamma/hadron separation cuts have been applied.   }
\label{fig:diff}
\end{figure}

Figure \ref{fig:diff} shows the difference between the two energy estimators as a function of $E$. A systematic difference can be seen at low energies, with the NN returning, on average, a lower estimate than the GP. At high energies, there is almost no systematic difference.

Two quantities are used to evaluate the energy-dependent performance of the estimators. The first is the resolution: the standard deviation of the energy estimate in log-energy space. The second is the bias, defined as the average difference between the reconstructed and true energies in log space:

\begin{equation}
b \equiv \left\langle\log_{10} \hat{E} - \log_{10} E\right\rangle.
\end{equation}

\begin{figure}
\includegraphics[width=0.48\textwidth]{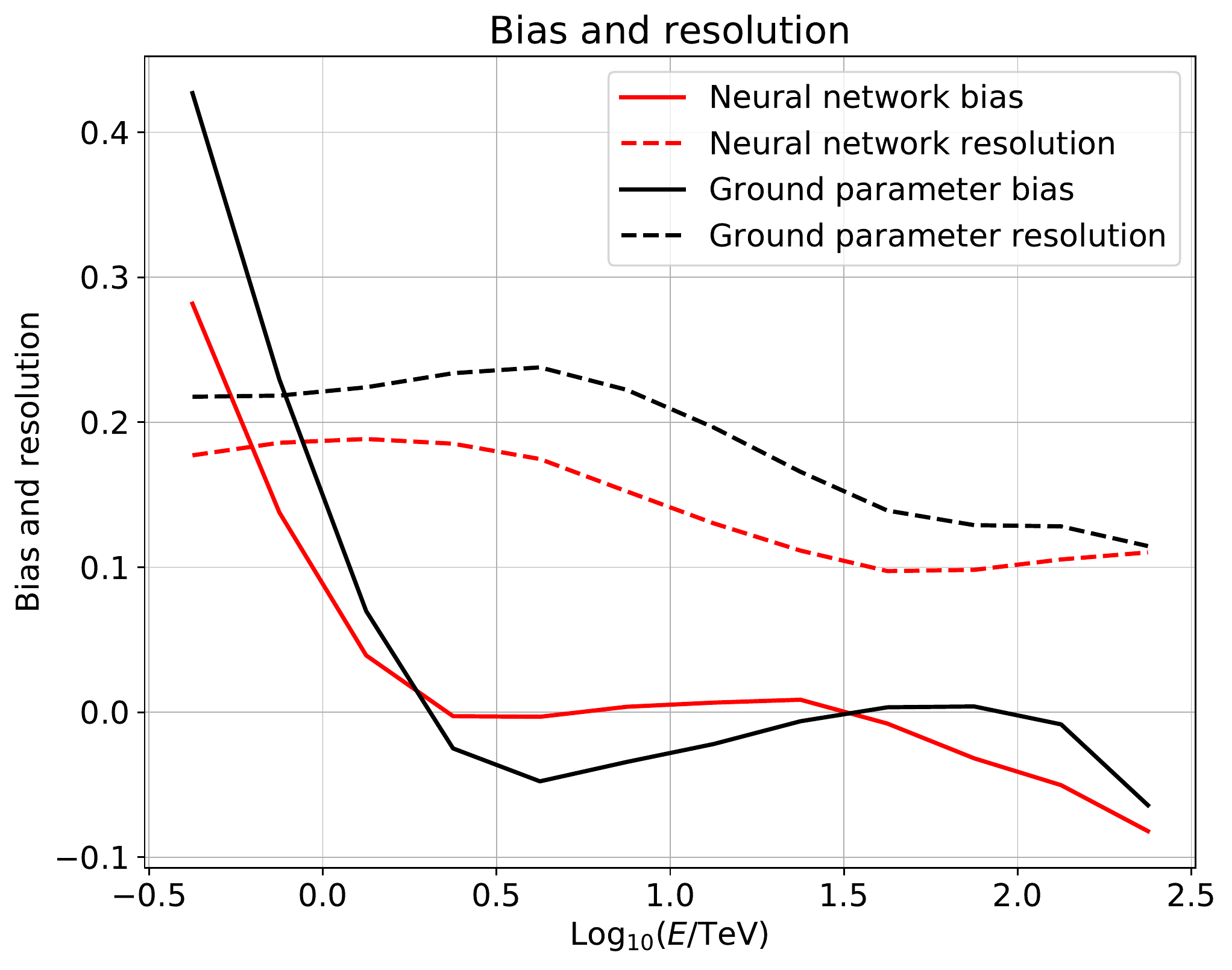}
\caption{The bias and resolution for both energy estimates. Bias is defined as the average difference between the reconstructed and true energies in log$_{10}$ space. Resolution is defined as the standard deviation of the energy estimate, also in log$_{10}$ space. Gamma/hadron separation cuts have been applied. The large bias at the lowest energies is because of the event selection requirement that a minimum number of PMTs be hit, which leaves only air showers with upward fluctuations in the number of PMTs hit.}
\label{fig:bias}
\end{figure}

The bias and resolution for both estimators can be seen in Figure \ref{fig:bias}. Both the NN and GP have a large bias below 1 TeV.  This is due to the event selection requirement that a minimum of 6.7\% of the array be hit, which remove the vast majority of events below this energy. The only events left are from air showers with upward fluctuations in the number of PMTs hit. Due to the substantial bias and poor resolution below 1 TeV, events with reconstructed energies below this threshold are excluded from the spectral fit. HAWC is not as sensitive to GeV energies as it is to TeV energies, so this choice does not affect the fit. 

Note that both estimators have very good resolution (less than the bin width of log$_{10}$(E/TeV) = 0.25) and almost no bias in the high-energy regime (between 10 and 316 TeV). The NN has a more favorable bias below $\sim$32 TeV, while the ground parameter performs better above this energy.

\begin{figure}
\includegraphics[width=0.48\textwidth]{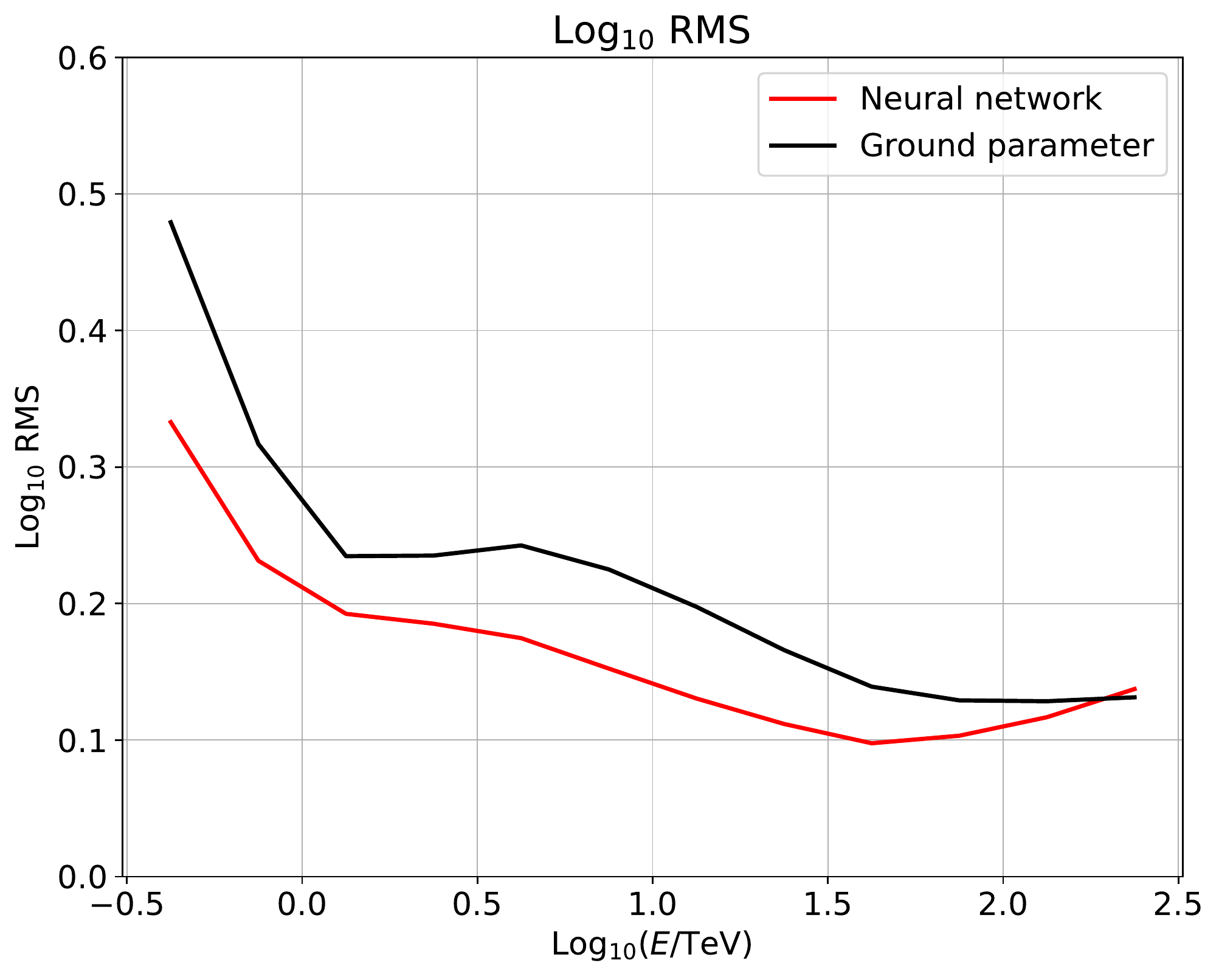}
\caption{The log RMS error for the GP and NN estimators. Gamma/hadron separation cuts have been applied.  This is defined as $\rho \equiv \sqrt{\left\langle\left(\log_{10} \hat{E} - \log_{10}E\right)^2 \right\rangle}.$}
\label{fig:rmserror}
\end{figure}

The log RMS error is defined as

\begin{equation}
\rho \equiv \sqrt{\left\langle\left(\log_{10} \hat{E} - \log_{10}E\right)^2 \right\rangle}.
\end{equation}
This is the bias and resolution added in quadrature. Figure \ref{fig:rmserror} shows the log RMS error for both energy estimators. The NN performs better using this metric.

\section{Measurement of the very high energy Crab spectrum}\label{sec:crab}

\subsection{Dataset}

The data used in this analysis were collected between June 2015 and December 2017. The total livetime is $\sim$837 days. The detector had $> 90\%$ uptime during this period. The loss of livetime comes from days where the detector was off for maintenance or due to operational difficulties. Additionally, a small amount of data were removed due to large variances in the zenith angle distribution, which is an indication that the detector was unstable during that period. These instabilities make the background estimation method unusable, so such data are removed. This background estimation technique is described in Section \ref{sec:bkg}.

\subsection{Event selection and binning}

The spectral fit is performed using a binned-likelihood technique. This forward-folding method accounts for bias and resolution in the energy estimate. We use a 2D binning scheme based on $\mathcal{B}$ (described in Section \ref{sec:hawc}) and the estimated energy. This 2D binning scheme was chosen instead of binning solely in energy because the gamma/hadron separation parameters as well as the angular resolution depend on both the energy and size of the event.  We use 9 $\mathcal{B}$ bins each subdivided into 12 energy bins, for a total of 108 bins.  The energy bins are quarter-decade bins in log$_{10}(\hat{E})$, beginning at log$_{10}(\hat{E}/\mathrm{TeV})$ = -0.5 (\replaced{316 GeV}{.316 TeV}) and ending at log$_{10}(\hat{E}) = 2.5$ (316 TeV). See Tables  \ref{table:bins} and \ref{table:fhitbins} for the bin definitions. For example, bin 9k would be all of the events with $> 84\%$ of the array hit and energies between 100 TeV and 177~TeV.

In practice, not all 108 bins are used. Some of these bins have little to no probability that events will populate them; for example, there are no low-energy events where the entire array is hit. Additionally, some of these bins contain so few events that they are not modeled well in the Monte Carlo simulation and are also excluded from the fit. \added{The bins used in the fit were chosen \textit{a priori} by looking at the distribution of estimated energies across each simulated $\mathcal{B}$ bin and keeping the central 99$\%$ of the events. This removes empty bins as well as the tails of the distribution, where statistics are low and there are more likely to be mismodeled events and a data/Monte Carlo simulation discrepancy.} The effect of this exclusion is discussed in the systematic uncertainty section (Section \ref{sec:sys}). In this analysis, 40 bins are used in the ground parameter fit and 37 bins in the neural network fit.

\begin{table}
\begin{center}
\caption{Energy bins}
\begin{tabular}{|c|c|c|}
\hline
Bin & Low energy (TeV) & High energy (TeV)\\
\hline
a &   0.316 &   0.562 \\
b &   0.562 &   1.00  \\
c &   1.00  &   1.78  \\
d &   1.78  &   3.16  \\
e &   3.16  &   5.62  \\
f &   5.62  &  10.0   \\
g &  10.0   &  17.8   \\
h &  17.8   &  31.6   \\
i &  31.6   &  56.2   \\
j &  56.2   & 100     \\
k & 100     & 177     \\
l & 177     & 316     \\
\hline
\end{tabular} \end{center}

The energy bins. Each bin spans one quarter decade. Note that the first two bins are not used in this analysis as the estimate is highly biased, as explained in Section \ref{sec:performance}.
\label{table:bins}
\end{table}

\begin{table}
\begin{center}
\caption{$\mathcal{B}$ bins}
\begin{tabular}{|c|c|c|}
\hline
Bin number & Low fraction hit & High fraction hit\\
\hline
1 & 0.067 & 0.105 \\
2 & 0.105 & 0.162 \\
3 & 0.162 & 0.247 \\
4 & 0.247 & 0.356 \\
5 & 0.356 & 0.485 \\
6 & 0.485 & 0.618 \\
7 & 0.618 & 0.740 \\
8 & 0.740 & 0.840 \\
9 & 0.840 & 1.00  \\
\hline
\end{tabular}
\end{center}

The $\mathcal{B}$ (fraction of PMTs hit) analysis bins used in this paper.
\label{table:fhitbins}
\end{table}

Several improvements have been made from \cite{Abeysekara2017} to strengthen the analysis.  A requirement that the shower core be reconstructed on the HAWC array has been added. This improves the angular and energy resolutions, as events with cores on the array are typically better reconstructed.

Gamma/hadron separation has also been improved (through refinements to the simulation that have improved the data/Monte Carlo simulation agreement), although the gamma/hadron separation variables used in this analysis are unchanged from \cite{Abeysekara2017}. Compactness, first described in \cite{Abeysekara2013}, is effective at identifying air showers containing muons. Muons, dominantly present in hadronic (background) showers, appear as localized charge depositions far from the shower core. The second parameter is known as PINCness \citep{Abeysekara2017} and measures the smoothness of the LDF. Gamma-ray showers have smoother profiles than hadronic air showers.

Although the gamma/hadron separation variables are unchanged, the actual cut values have been reoptimized in each 2D $\mathcal{B}$/energy bin. This allows for better identification of the highest-energy events as compared to \cite{Abeysekara2017}, where nearly everything above 30 TeV was included in one analysis bin and had the same gamma/hadron cuts.  The cut values are determined \textit{a priori} using simulated Crab signals and background data, with a requirement that each bin has at least 50\% gamma-ray efficiency. The efficiency to gamma rays in a given bin ranges from 50\% to nearly 100\%. The gamma/hadron cuts are optimized separately for each estimator.  Additionally, the data quality cuts described in Section \ref{sec:biasres} have also been applied to the data.

\begin{figure*}
\centering
\includegraphics[width=0.8\textwidth]{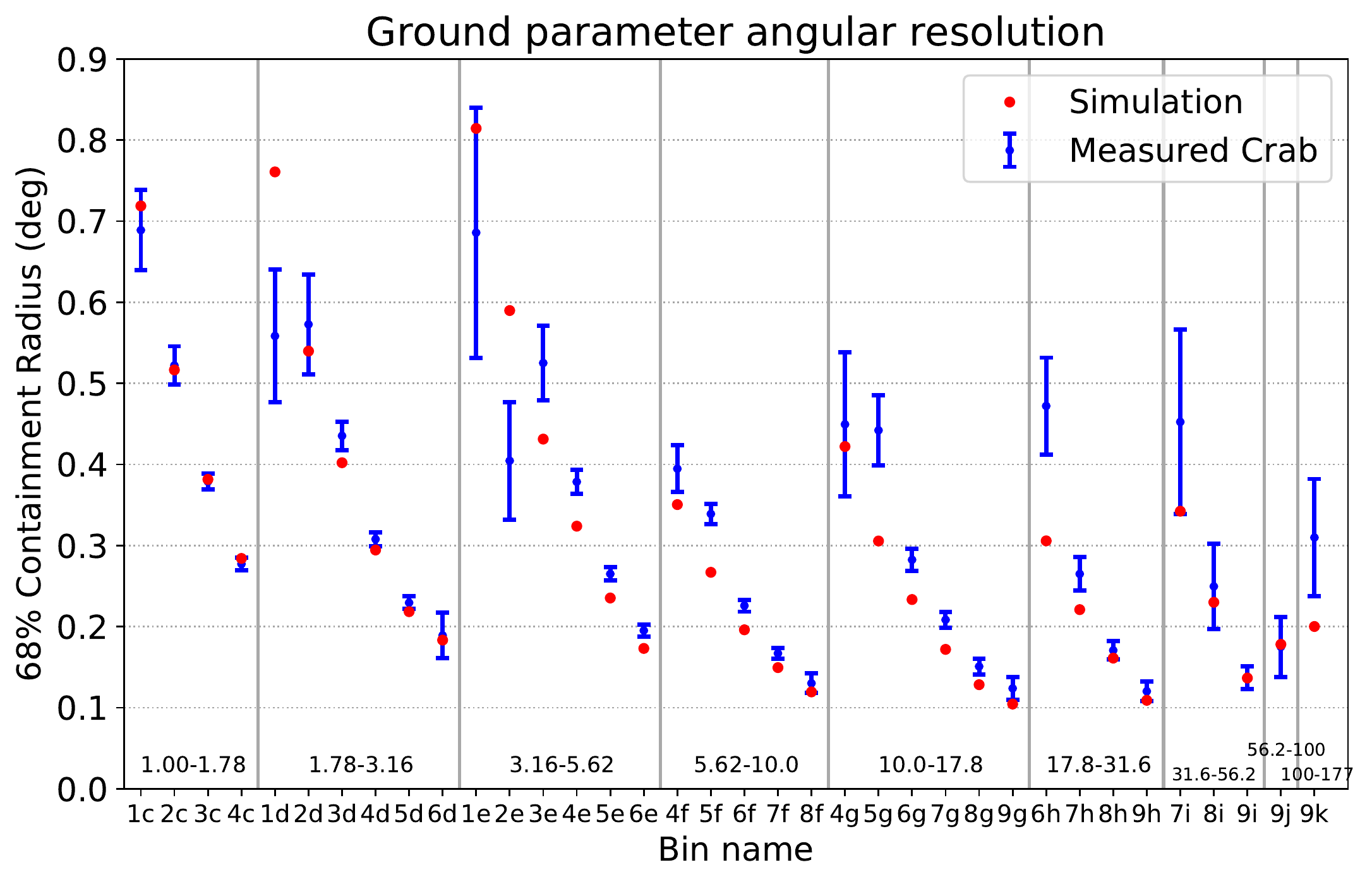}
\subfloat{\includegraphics[width=0.8\textwidth]{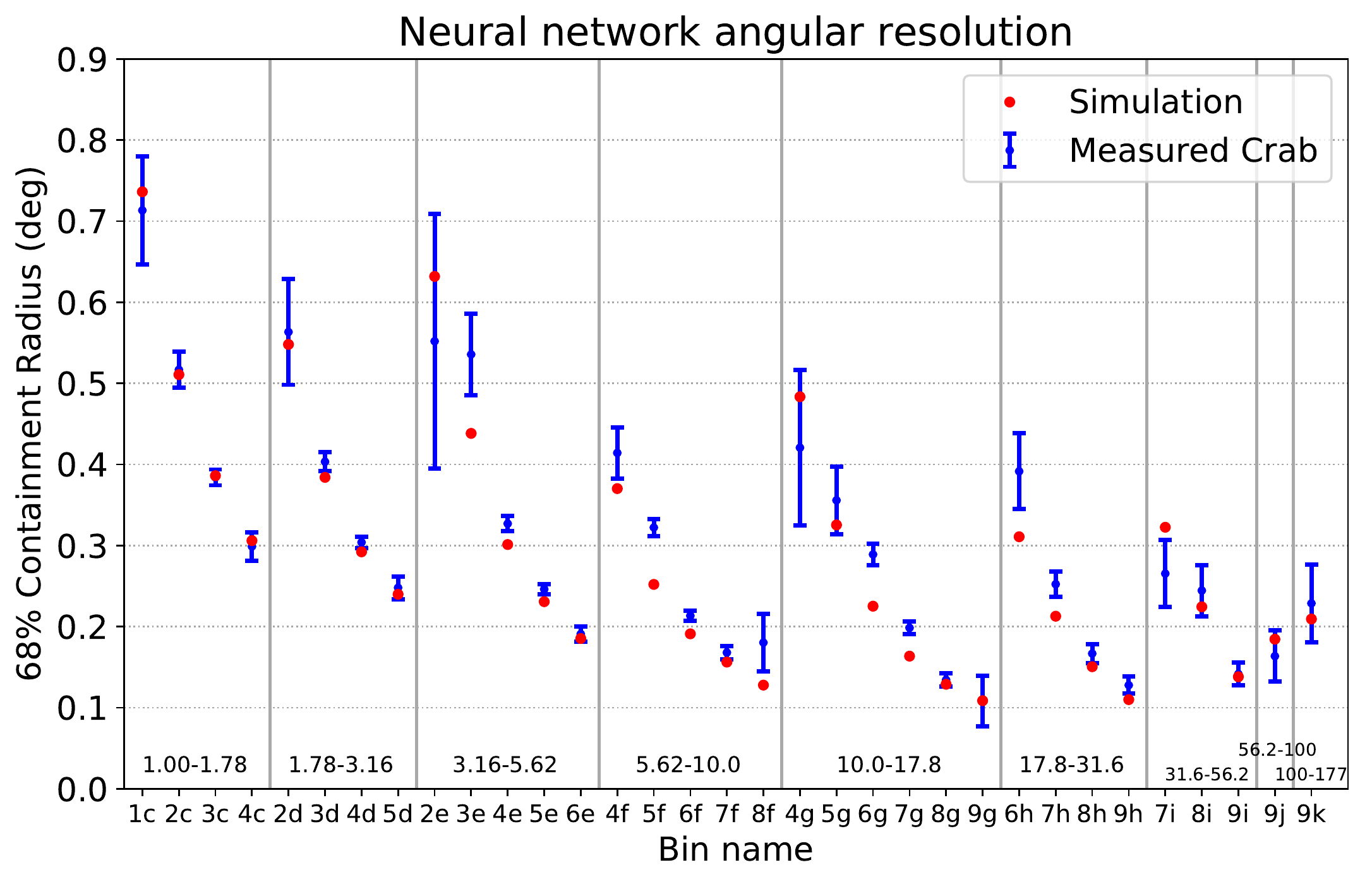}}

\caption{The 68\% containment values in data and Monte Carlo simulation for the ground parameter energy estimator (top) and neural network (bottom). Only bins where the Crab Nebula is detected $> 3\sigma$ are shown.  The plot is arranged so that bins contributing to a given energy bin are collected together in order of increasing $\mathcal{B}$ value, with divisions between estimated energy bins given by the vertical grey lines. The reconstructed energy ranges are labeled. The data/MC discrepancy visible in the figure is small ($\sim$5$\%$) and treated in the systematic uncertainty analysis. It is a subdominant contribution to the overall systematic uncertainty. This is discussed further in Section \ref{sec:angresdis}.}
\label{fig:angres}
\end{figure*}

Another improvement is the point spread function (PSF). As before, this is modeled as a linear combination of two two-dimensional Gaussians, determined from simulated events. Better modeling of this PSF is one of the significant changes from the previous Crab analysis. The radius required to contain 68$\%$ of the photons has a strong dependence on both the event size and energy, so the 2D binning scheme used here allows for a more precise determination of the PSF (see Figure \ref{fig:angres}).  For example, all events from $\mathcal{B}$ bin 1 in \cite{Abeysekara2017} had a 68$\%$ containment of $\sim$1 degree. Here, these events have a 68$\%$ containment between $\sim$0.27$^{\circ}$ and $\sim$0.75$^{\circ}$, depending on the energy of the shower.

Lastly, note that the definition of $\mathcal{B}$ has changed slightly from \cite{Abeysekara2017}: there, $\mathcal{B}$ was defined as the number of PMTs detecting light divided by the total number of PMTs that were operational at the time. Here, the numerator is changed to the fraction of PMTs detecting light within 20 ns of the shower front. This change reduces the number of noise hits contributing to the size of the event.

\subsection{Background estimation}
\label{sec:bkg}
For small showers, hadronic cosmic rays dominate over gamma rays even after gamma/hadron separation cuts have been applied. An estimate of this cosmic-ray background is performed individually in each analysis bin. For the lower energy bins, where there are many events, the standard HAWC background estimation technique is applied. This is known as ``direct integration.'' This algorithm was originally developed by the Milagro Collaboration \citep{Atkins2003} and has become the standard HAWC background estimation algorithm. As described in \cite{Abeysekara2017}, the background estimate from direct integration is smoothed by an additional 0.5$^{\circ}$ to compensate for the sparseness of the background.

In the highest-energy bins, the statistics are too low to give a spatially smooth background estimate using the 2 hour chunks of data that are the backbone of direct integration. A different algorithm known as ``background randomization'', similar to the one in \cite{Alexandreas1991}, is used to average over the entire dataset and give a spatially smooth background estimate for these low-background bins. For each bin where the all-sky rate is less than 500 events per day, a 2D distribution of the local coordinates (zenith and azimuth) is constructed. A random (zenith, azimuth) pair is drawn from this distribution for each event and used with the time of the event to calculate a right ascension and declination, which is added to the background map. This process is repeated 10,000 times for each event; the background map is then normalized to the number of events in the map. This produces a background estimate much smoother than given by direct integration. Direct integration is still used for higher-statistics bins, as it is less computationally intensive and is needed to correctly incorporate the cosmic ray anisotropy into the background estimate. 

The background estimation technique described above has the potential to be systematically biased if the local coordinate distributions are not stable in time.  The zenith and azimuthal angle distributions have been checked and found to have the required stability.

\subsection{Likelihood fit}

The functional form assumed for the forward-folded fit is a log-parabola:

\begin{equation}
\label{eq:log parabola}
\frac{dN}{dE} = \phi_0 \left(E/E_0\right)^{-\alpha-\beta\ln\left(E/E_0\right)}.
\end{equation}
Previous measurements indicate that a log parabola is likely to be a good fit to the Crab Nebula spectrum. The pivot energy, $E_0$, was chosen to be 7 TeV to minimize correlations with the other parameters. The other parameters are free in the fit, which is performed using the HAWC plug-in to the Multi-Mission Maximum Likelihood framework \citep{Younk2015, Vianello2015}, an analysis pipeline that is capable of handling data from a wide variety of astrophysical detectors. The spectral parameters $\phi_0$, $\alpha$, and $\beta$ are chosen to maximize the test statistic

\begin{equation}
\label{eq:ts}
TS \equiv 2 \ln \frac{L_\text{S+B}\!\left(\phi_0, \alpha, \beta\right)}{L_\text{B}},
\end{equation}
where $L_\text{S+B}$ is the likelihood for the signal-plus-background hypothesis and $L_\text{B}$ is the likelihood for the background-only hypothesis.

Although the Crab Nebula is slightly extended at TeV energies \citep{Holler2017}, it is modeled as a point source here.  HAWC lacks the angular resolution to measure the extent. 

\begin{figure*}
\centering
\includegraphics[width=0.8\textwidth]{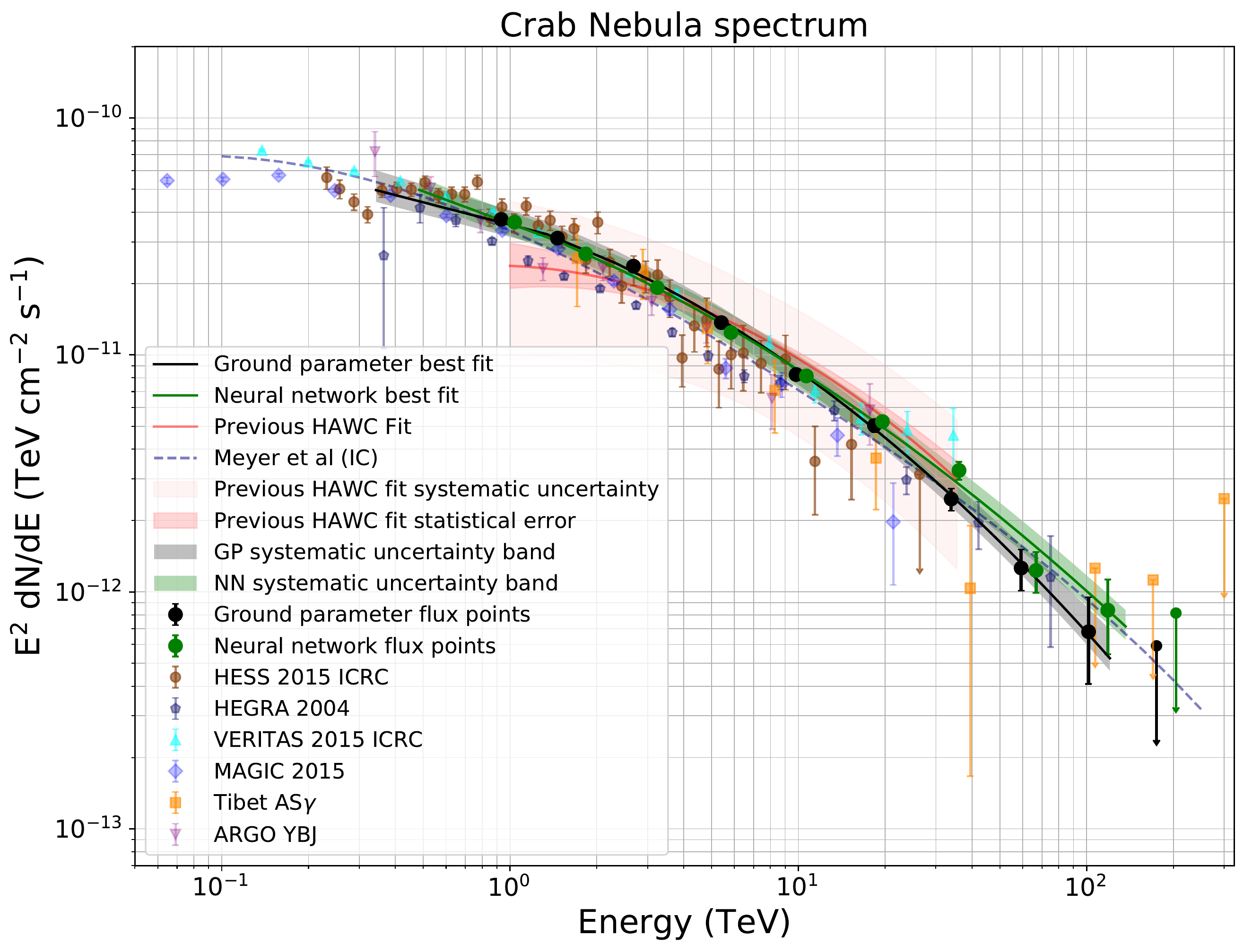}
\caption{The Crab spectrum obtained with the GP method (black) and NN method (green). The error bars on the flux points are statistical only The shaded grey and green shaded bands denote systematic uncertainties. The upper ranges of the overall forward-folded fit are calculated using binomial statistics (described in Section \ref{sec:detection}). This method breaks down when there are large numbers of events, so the lower ranges of the fits are chosen by looking at the simulated energy distribution in the lowest-energy bin and finding the energy that 90$\%$ of the events in that bin are above.  For comparison, the HAWC Crab fit from \cite{Abeysekara2017} is also shown. See the text for details of how the flux points were obtained. Systematic uncertainties are discussed further, in Section \ref{sec:sys}. The dotted navy line is the Inverse Compton parameterization from \cite{Meyer2010}. References for other experiments: HESS~\citep{Holler2015}, VERITAS~\citep{Meagher2015}, MAGIC~\citep{Aleksic2015}, Tibet AS$\gamma$~\citep{Amenomori2015}, ARGO YBJ~\citep{Bartoli2015}, HEGRA~\citep{Aharonian2004}}
\label{fig:spectrum}
\end{figure*}

The spectra of the Crab Nebula obtained using the two energy estimators can be seen in Figure \ref{fig:spectrum}, and the global best-fit parameters over the HAWC energy range can be seen in Table \ref{table:results}. Uncertainties quoted in the table are statistical only. Systematic uncertainties are discussed in Section \ref{sec:sys}.  The test statistic is 17995 for the ground parameter and 19402 for the neural network. The chi square per degree of freedom ($\chi^2$/NDF) is approximately 1.7 for both the GP and NN log parabola fits, dominated by the low energy (high statistics) bins. Adding a systematic uncertainty of 1-2$\%$ in quadrature with statistical uncertainties reduces the $\chi^2$/NDF to 1. This value was computed using 2 times the optimal tophat radius in each 2D bin.

Alternative spectral models were also considered.  For a power law, the test statistic is 17865 (19347) for the GP (NN). For a power law with an exponential cutoff, the test statistic is 17979 (19395) for the GP (NN). We report the log parabola as the nominal spectrum because it offers the most improvement over a power law for both energy estimation methods over the HAWC energy range.

\begin{table}
\begin{center}
\caption{Likelihood fit results}
\begin{tabular}{|c||c|c|c|}
\hline
Estimator & $\phi_0$ & $\alpha$ & $\beta$ \\
 & ($10^{-13}$ TeV cm$^2$ s)$^{-1}$ & & \\
\hline\hline
GP & 2.35$\pm$0.04 & 2.79$\pm$0.02 & 0.10$\pm$0.01 \\
NN & 2.31$\pm$0.02 & 2.73$\pm$0.02 & 0.06$\pm$0.01\\
\hline
\end{tabular} \end{center}

The results of the likelihood fit to a log-parabola shape for each estimator. Uncertainties are statistical only. 
\label{table:results}
\end{table}

\begin{table*}
\begin{center}
\caption{Test statistic as a function of energy and flux points}
\begin{tabular}{|c||c||c|c|c||c|c|c|}
\hline
Bin & $\hat{E}$ energy range & GP TS & GP median & GP flux & NN TS & NN median & NN flux \\
& (TeV) & & energy (TeV) & (TeV cm$^{-2}$ s$^{-1}$) & & energy (TeV) &(TeV cm$^{-2}$ s$^{-1}$) \\
\hline\hline
c & 1-1.78 & 3896 & 0.932 & (3.73 $\pm$ 0.07) $\times$ 10$^{-11}$ & 2734 & 1.04 & (3.63 $\pm$ 0.08) $\times$ 10$^{-11}$ \\
d & 1.78-3.16 & 3754 & 1.46 & (3.11 $\pm$ 0.07) $\times$ 10$^{-11}$ & 4112 & 1.83 & (2.67 $\pm$ 0.05) $\times$ 10$^{-11}$ \\
e & 3.16-5.62 & 3543 & 2.68 & (2.37 $\pm$ 0.06) $\times$ 10$^{-11}$ & 4678 & 3.24 & (1.92 $\pm$ 0.04) $\times$ 10$^{-11}$ \\
f & 5.62-10.0 & 3481 & 5.41 & (1.37 $\pm$ 0.04) $\times$ 10$^{-11}$ & 3683 & 5.84 & (1.24 $\pm$ 0.03) $\times$ 10$^{-11}$ \\
g & 10.0-17.8 & 1864 & 9.82 & (8.26 $\pm$ 0.33) $\times$ 10$^{-12}$ & 2259 & 10.66 & (8.15 $\pm$ 0.31) $\times$ 10$^{-12}$\\
h & 17.8-31.6 & 975 & 18.4 & (5.04 $\pm$ 0.31) $\times$ 10$^{-12}$ & 1237 & 19.6 & (5.23 $\pm$ 0.29) $\times$ 10$^{-12}$ \\
i & 31.6-56.2 & 365 & 33.9 & (2.47 $\pm$ 0.27) $\times$ 10$^{-12}$ & 572 & 36.1 & (3.26 $\pm$ 0.28) $\times$ 10$^{-12}$ \\
j & 56.2-100 & 107 & 59.3 & (1.26 $\pm$ 0.25) $\times$ 10$^{-12}$ & 105 & 66.8 & (1.23 $\pm$ 0.24) $\times$ 10$^{-12}$ \\
k & 100-177 & 19.9 & 102 & (6.79 $\pm$ 2.70) $\times$ 10$^{-13}$ & 28.8 & 118 & (8.37 $\pm$ 2.91) $\times$ 10$^{-13}$ \\
l & 177-316 & 0.33 & 174 & $<$ 5.92 $\times$ 10$^{-13}$ & 0.14 & 204 & $<$ 8.14 $\times$ 10$^{-13}$\\
\hline
\end{tabular}
\\
The test statistic for each energy bin, corresponding to the flux points in Figure \ref{fig:spectrum}. The ``$\hat{E}$ energy range'' column gives the range in reconstructed energy for each bin, while the columns labeled ``GP med. energy'' and ``NN med. energy" give the median energy from simulation for the ground parameter and neural network, respectively, assuming that the fitted log parabola spectra are the true spectra. Some median energies fall outside the reconstructed energy range because the Crab Nebula spectrum is steep, so that there are more photons with lower energy than higher which are reconstructed at a given $\hat{E}$. The flux gives statistical uncertainties only and is reported at the median energy in each bin. The last bin is a 95$\%$ upper limit following \cite{Feldman1998}. 
\label{table:ts}
 \end{center}
\end{table*}

Flux points are calculated by holding $\alpha$ and $\beta$ constant from the global fit and fitting the normalization ($\phi_0$) individually for each group of $\mathcal{B}$ bins that contribute to a given reconstructed energy bin, similar to \cite{Yuan:2013qwa}. While this is not a full unfolding prescription, it allows one to see if any energy bins are inconsistent with the fitted log-parabola spectrum. Figure \ref{fig:spectrum} shows the Crab Nebula spectrum computed in this manner. The error bars are statistical only. Systematic uncertainties (discussed in Section \ref{sec:sys}) are shown as a band over the global forward-folded fit. Points are shown for each reconstructed energy bin where the statistical significance in the fit is above 2$\sigma$.  The flux points are plotted at the median \replaced{true}{simulated} energy in each bin, as determined from the Monte Carlo simulation.

The TS for each flux point in each of the two spectra (Figure \ref{fig:spectrum}) are listed in Table \ref{table:ts}. Since the test statistic in the last bin ($> 177$ TeV) is only 0.33 for the GP and 0.14 for the NN, upper limits are set in this bin using a 95\% upper confidence interval following \cite{Feldman1998}.

\begin{figure}
\includegraphics[width=0.48\textwidth]{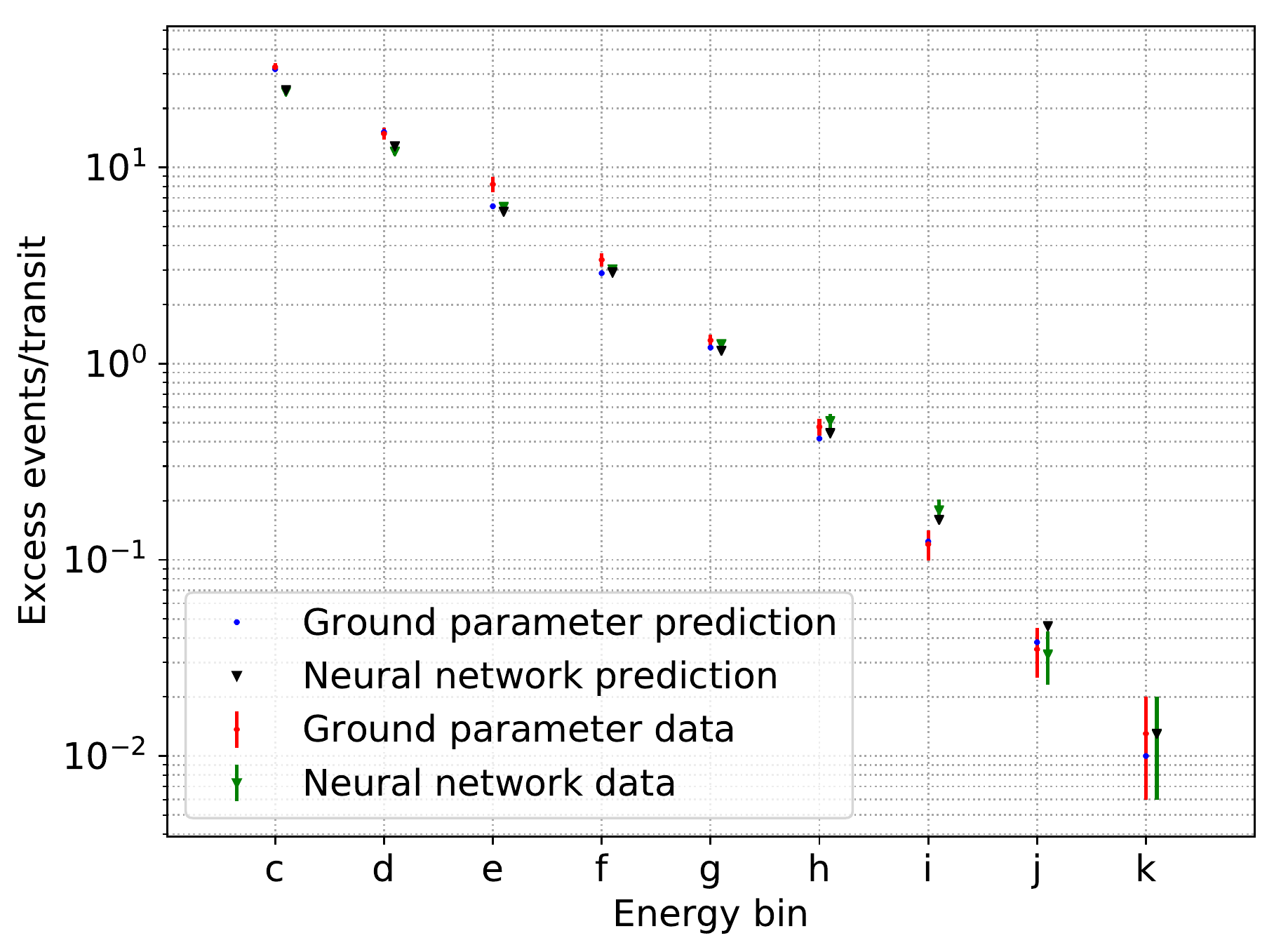}
\subfloat{\includegraphics[width=0.48\textwidth]{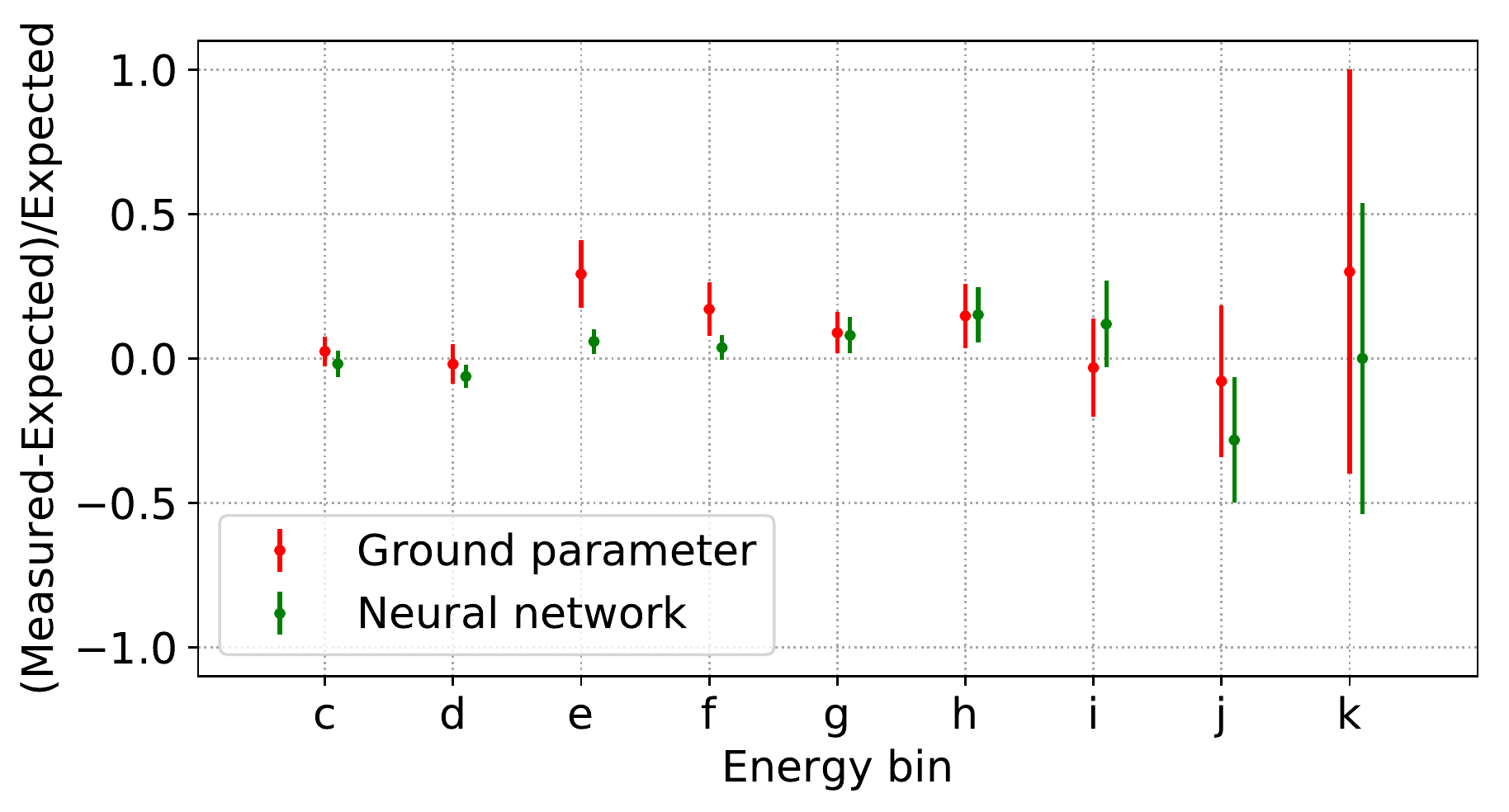}}
\caption{The Crab excess per transit, along with the residual, defined as (measured - expected)/expected. The two estimators have different numbers of events in some bins due to differing bias, resolution and efficiency to gamma rays.}
\label{fig:excess}
\end{figure}

The measured excess per transit, along with the expected value from simulation, can be seen in Figure \ref{fig:excess}. Assuming that the true spectrum is the measured HAWC spectrum, the simulation predicts 57.87 gamma rays with a reconstructed energy above 1 TeV per day from the Crab Nebula using the GP analysis chain and 48.36 using the NN analysis chain. The values are different because the gamma/hadron cuts were optimized separately for the two techniques and they therefore have different efficiencies to gamma rays. In the data, we observe an excess of 60.85 $\pm$ 2.10 gamma rays with the ground parameter and 47.72 $\pm$ 1.28 with the neural network, consistent with expectations. All values are computed using a 2 degree radius centered on the Crab Nebula location.

\subsubsection{Bin contamination in spectral fits}
Bin purity measures the contamination of a reconstructed-energy bin by mis-reconstructed events. It is defined here as the fraction of events in a quarter-decade reconstructed energy bin whose \replaced{Monte Carlo true}{simulated} energy is also within that bin:

\begin{equation}
p \equiv P\!\left(E \in B\middle|\hat{E} \in B\right),
\end{equation}
where $B$ is a quarter-decade energy bin. Both bias and energy resolution can affect the bin purity. 
 
\begin{figure}
\includegraphics[width=0.48\textwidth]{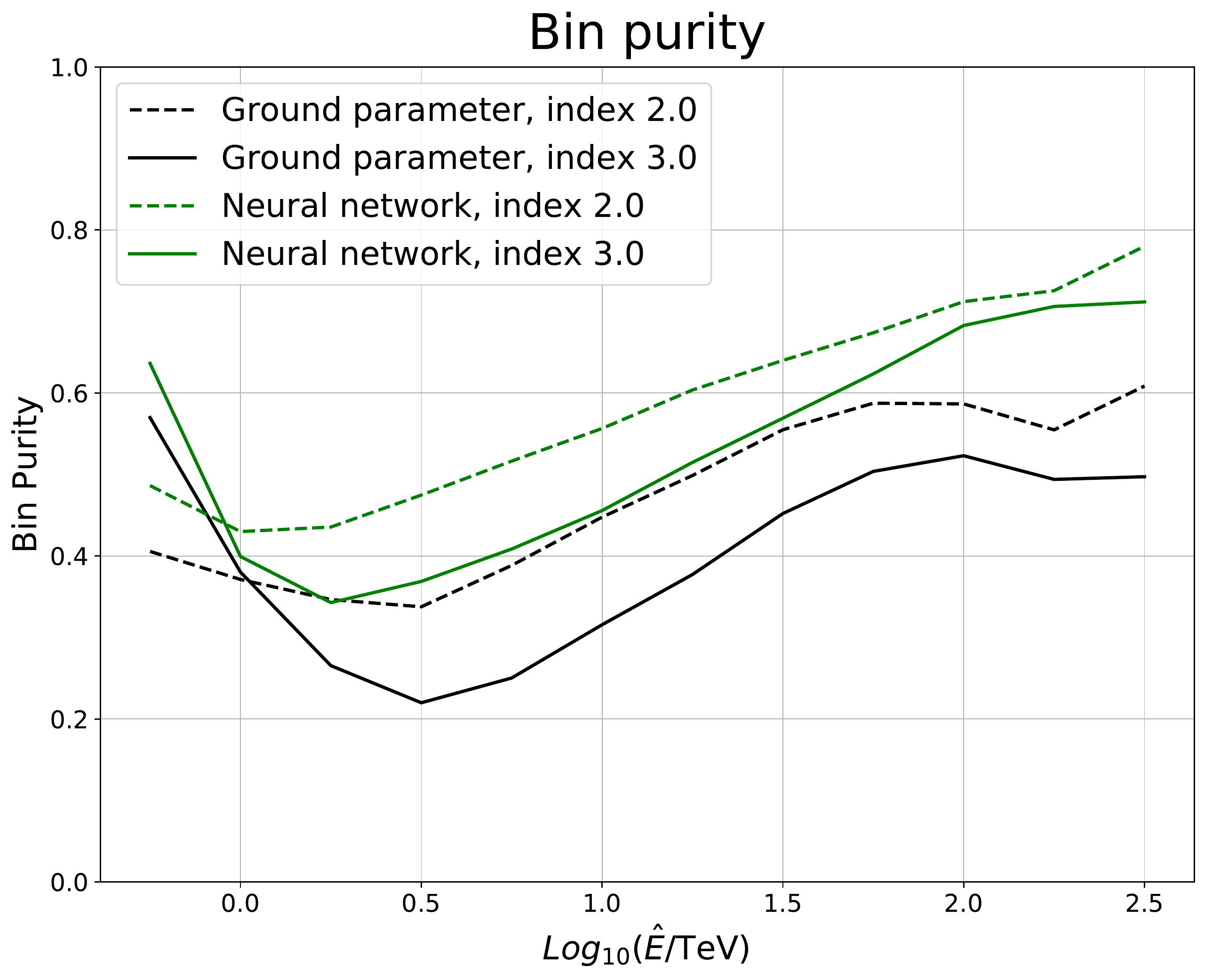}
\caption{The bin purity for both estimators, for a hard (E$^{-2}$) and a soft (E$^{-3}$) power-law spectra. The plot is made after gamma/hadron cuts.}
\label{fig:purity}
\end{figure}

Because astrophysical sources emit following roughly power-law spectra with negative spectral indices, there are many more lower-energy gamma rays than higher energy ones. If even a small percentage of these low energy gamma rays are mis-reconstructed with a higher energy, the bin purity can be adversely affected. Thus, this parameter is a function of spectral assumption. A softer spectrum will have worse bin purity. Figure \ref{fig:purity} shows the bin purity for both estimators. For a power-law spectrum with index between 2 and 3, bin purity is \replaced{$> \sim50\%$}{$\gtrsim$ 50$\%$} above 100 TeV. Bin purity can worsen if the observed gamma-ray spectrum has a cutoff or curvature.

Note that bin contamination is not a concern in the likelihood fit described above; since the fit is forward-folded, biases and energy resolution in the energy assignments are taken in account. However, bin purity is a concern if one wants to make a claim about the emission at the highest energies.

\subsubsection{Significance of the highest-energy detection}\label{sec:detection}
\begin{figure*}
\includegraphics[width=0.5\textwidth]{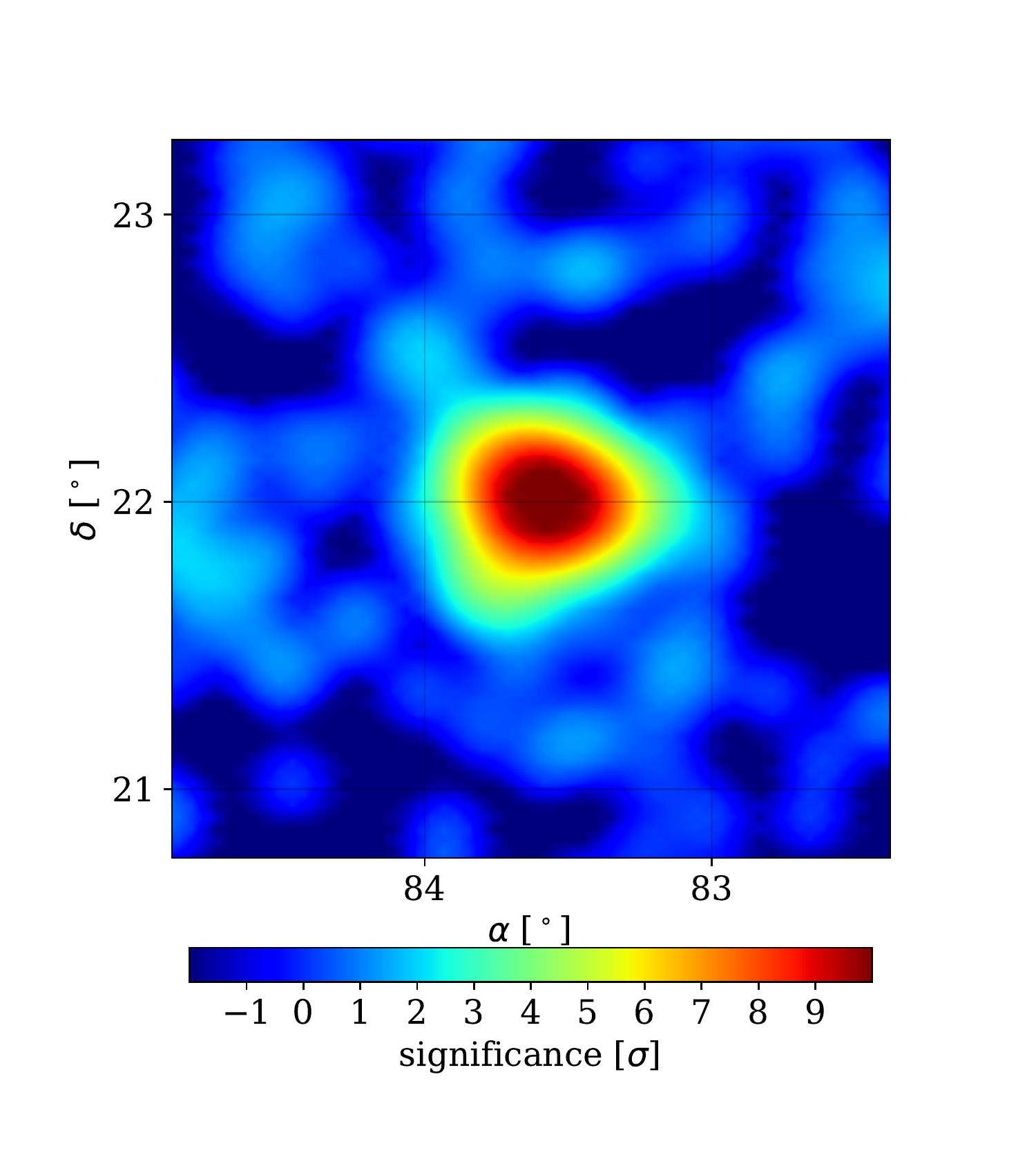}
\subfloat{\includegraphics[width=0.5\textwidth]{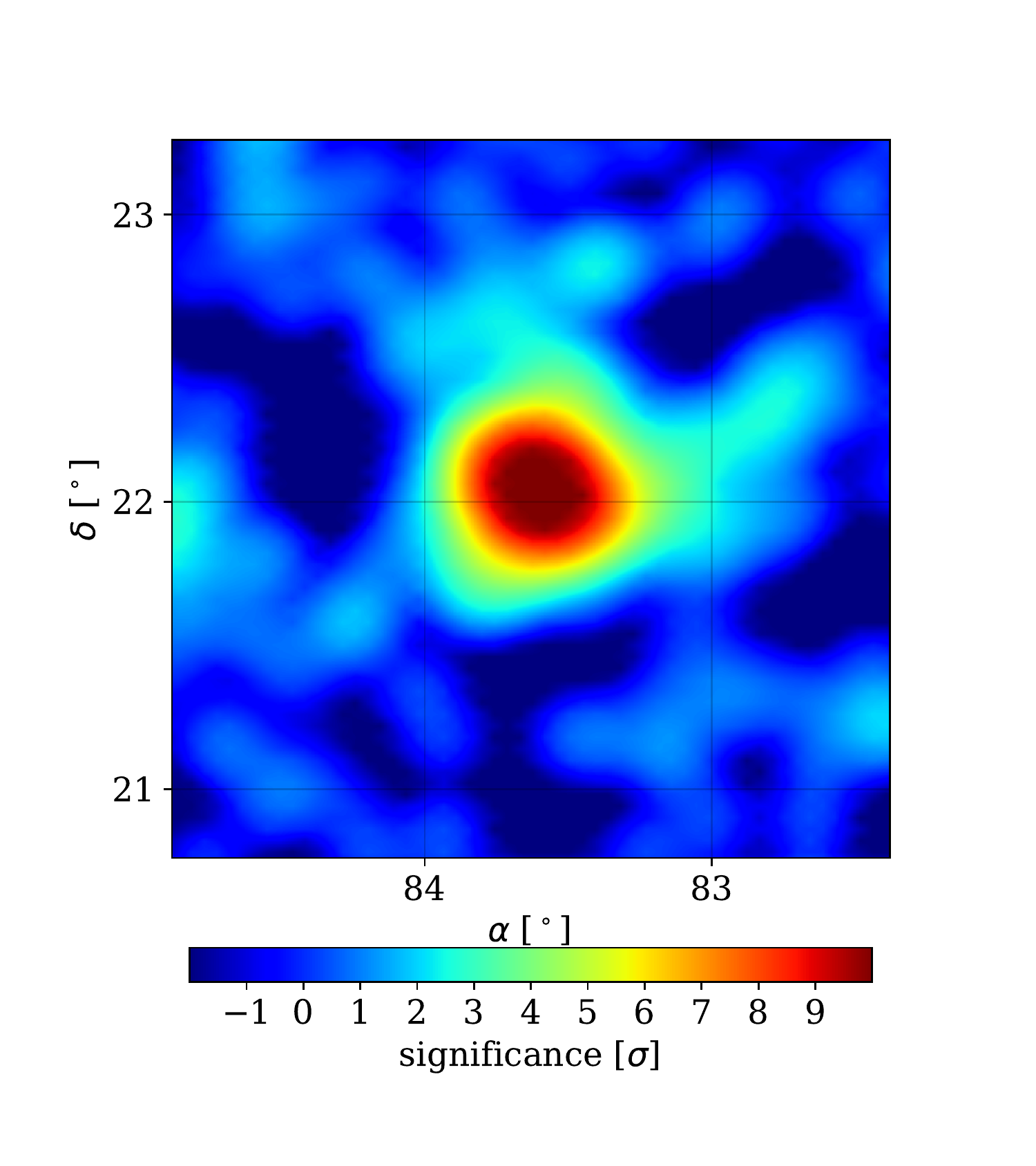}}
\caption{Significance map above 56 TeV in reconstructed energy for the ground parameter (left) and neural network (right). The maximum significance is 11.2$\sigma$ for the ground parameter and 11.6$\sigma$ for the neural network. Both significance maps have been smoothed for presentation purposes.  } 
\label{fig:gr56sig}
\end{figure*}

Detection of the Crab Nebula above $\sim$75 TeV would be the highest-energy detection of any astrophysical source to date. \added{Figure \ref{fig:gr56sig} gives a significance map of the region for $\hat{E} >$ 56 TeV for both estimators}. Figure \ref{fig:spectrum} provides flux estimates above 100 TeV. While at these energies the cosmic-ray background is significantly suppressed, the possibility of events that might have their energy overestimated and are statistical upward fluctuations from  lower energy bins becomes a concern. We investigate this possibility by fitting the Crab Nebula to the product of a log parabola and a step function, which effectively introduces the null hypothesis of no events above a certain hard-energy cutoff value. All of the parameters of the log parabola are left free.

We find that the conventional log parabola fit is significantly preferred over the log parabola convolved with a hard cutoff at 56 TeV for both estimators (5.12$\sigma$ for the GP and 6.99$\sigma$ for the NN, respectively). Moving the hard cutoff to 100 TeV, the conventional log parabola fit is preferred over the cutoff by 0.2$\sigma$ for the GP and 2.4$\sigma$ for the NN. The differences between the two methods can be explained by a combination of differences in the gamma/hadron cuts (which causes differences in gamma-ray efficiency), and statistical fluctuations. The neural network has a higher efficiency to gamma rays above 100 TeV. We interpret this as evidence for emission up to at least 100 TeV from the Crab Nebula. This forward folding procedure accounts for the energy resolution and bias, but ignores systematic uncertainties on the energy scale. This should be taken as a conservative approach to the maximum energy that emission from the Crab Nebula is detected at. 

Assuming that the measured energy spectrum of the Crab Nebula extends significantly past 100 TeV, we can use the procedure outlined in the following paragraphs to estimate the highest energy of the photons actually detected by HAWC.  

Table \ref{table:ts} gives the median energy from simulation for a source transiting at the Crab declination and with the best-fit spectra for each energy estimator. This number takes into account events that may have their energies over-estimated and are upward fluctuations from a lower-energy bin.  For both estimators, the last bin with a significant detection has a median energy above 100 TeV. The median energy is 102 TeV for the ground parameter and 118 TeV for the neural network. The somewhat large difference in median energies between the estimators can be explained by differing bin purities that stems from differences in energy resolution (see Figures \ref{fig:bias} and \ref{fig:purity}). This calculation assumes that the true spectrum of the Crab Nebula is the fitted log parabola.

We can expect roughly half of the $\sim$11 events in the 100-177 TeV bin to be above the median energy. From the binomial distribution, the probability of seeing zero events above the median is simply (0.5)$^{11}$, or 0.000488. This corresponds to a 3.3$\sigma$ detection of gamma rays above the median energy (102 TeV for the GP and 118 TeV for the NN).

If instead 2$\sigma$ is used as the threshold in the binomial calculation (which is the same threshold chosen for plotting flux points vs. setting an upper limit), the spectrum is detected to 121 TeV for the ground parameter and 137 TeV for the neural network. The spectra shown in Figure \ref{fig:spectrum} are plotted up to the 2$\sigma$ numbers from this binomial calculation.

\deleted{This will be further investigated in an upcoming publication on Lorentz invariance violation, which will discuss the highest-energy high-significance detection of the Crab Nebula. This publication will also include other high-energy emitting sources.}
\subsection{Systematic Uncertainties}\label{sec:sys}
The main sources of systematic uncertainties with HAWC comes from discrepancies between the data and the simulated Monte Carlo events that stem from uncertainties in the modeling of the detector. The systematic uncertainties described in section 4.3 of HAWC's previous Crab analysis \citep{Abeysekara2017} are present here, although improved detector modeling and constraints on the simulation parameters based on low-level data distributions have decreased the size of these uncertainties.  Rather than quoting one number for the systematic uncertainty on the flux, all of the uncertainties are treated in an energy-dependent manner for the first time.  This is an improvement over \cite{Abeysekara2017}, where the systematic uncertainty was quoted at $\pm$50$\%$ across HAWC's entire energy range.

We have looked for correlations between the sources of systematic uncertainty and have not found any. Therefore, the effect of each source of systematic uncertainty can be added in quadrature to the others.  The systematic uncertainties on each of the fit parameters in the log parabola likelihood fit can be seen in Table \ref{table:sysresults}. 

\begin{table}
\begin{center}
\caption{Systematic uncertainties on fit parameters}
\begin{tabular}{|c||c|c|c|}
\hline
 Estimator & Parameter & Sys. low & Sys. high \\
\hline\hline
GP & $\phi_0$ & -2.11$\times$10$^{-14}$ & 2.00$\times$10$^{-14}$ \\
 & $\alpha$ & -0.03 & 0.01 \\
 & $\beta$ & -0.03 & 0.01 \\
 \hline
NN & $\phi_0$ & -1.69$\times$10$^{-14}$ & 3.23$\times$10$^{-14}$ \\
 & $\alpha$ & -0.02 & 0.03 \\
 & $\beta$ & -0.02 & 0.02 \\
\hline
\end{tabular} \end{center}

The systematic uncertainties on the fit parameters, for each estimator. The units for $\phi_0$ are (TeV cm$^{-2}$ s$^{-1}$).
\label{table:sysresults}
\end{table}

\begin{figure*}
\includegraphics[width=0.48\textwidth]{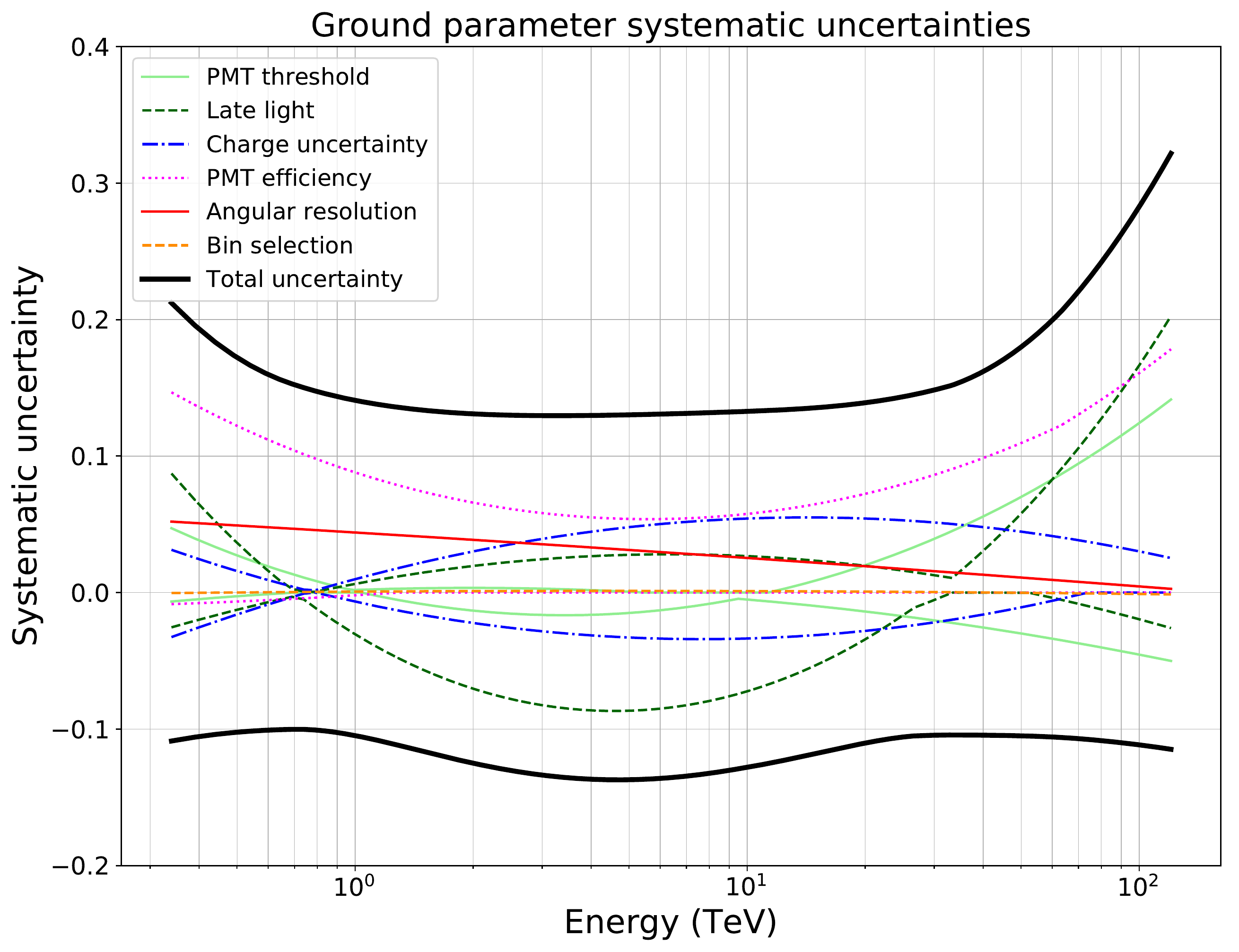}
\subfloat{\includegraphics[width=0.48\textwidth]{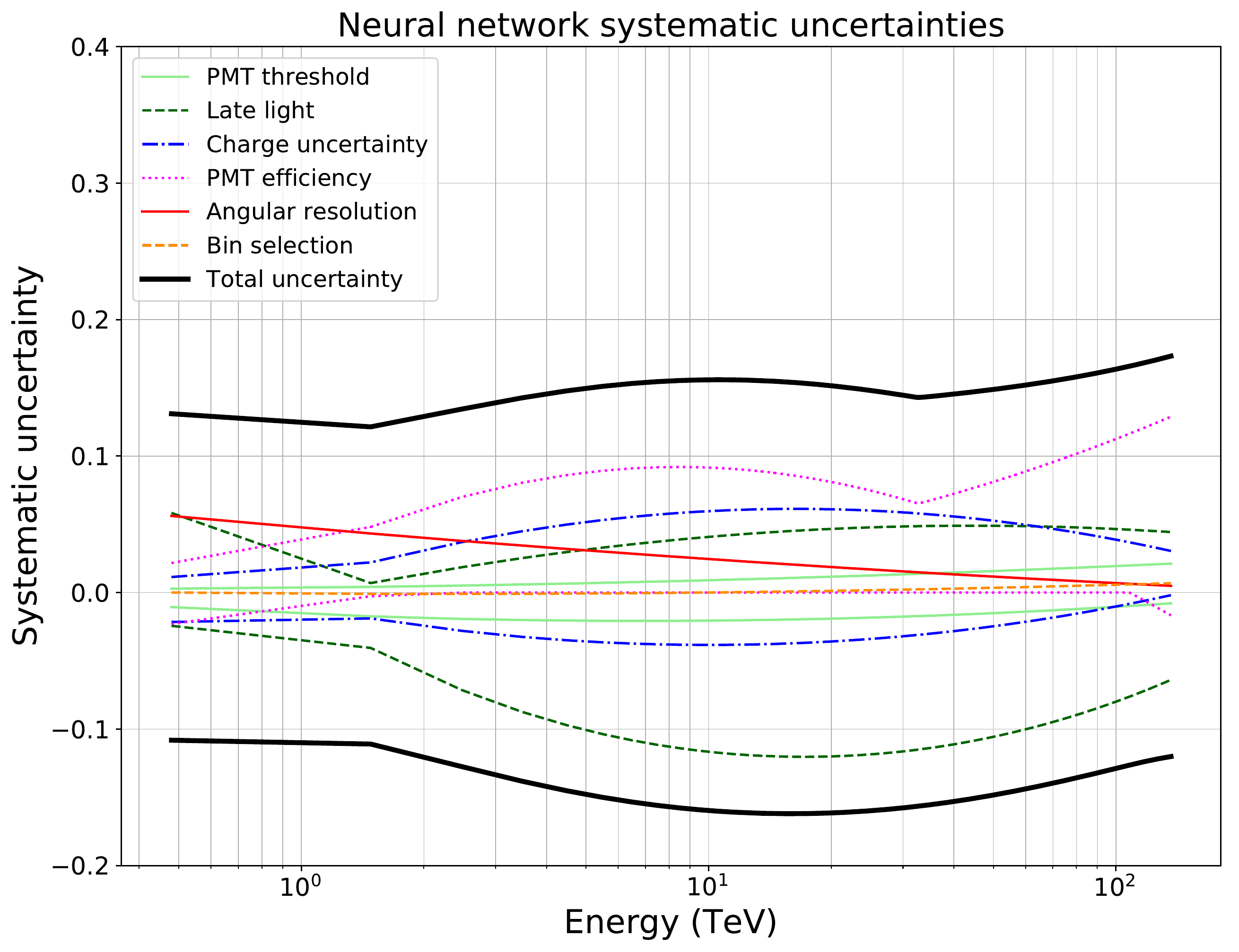}}
\caption{The contribution of each systematic uncertainty to the overall uncertainty in $E^2 dN/dE$, as a function of energy. Note that the y-axis scale is in linear space. The left figure shows the ground parameter and the right figure shows the neural network.  The thick black line is the total systematic uncertainty and includes an additional ten percent added in quadrature to conservatively cover systematic uncertainties not considered here (see Section \ref{sec:additional} for a discussion).}
\label{fig:sysboth}
\end{figure*}

The major sources of systematic uncertainty are described below.  Figure \ref{fig:sysboth} shows the shift due to systematics in $E^2 dN/dE$ as a function of energy for each estimator.

\subsubsection{Angular-resolution discrepancy}\label{sec:angresdis}
A discrepancy in the 68\% containment between data and simulation can be seen in Figure \ref{fig:angres}. While the cause of this is not immediately clear, it is thought to be at least partially because the shower curvature model used during reconstruction does not yet have an energy dependence.

The 68$\%$ containment in the Monte Carlo is underestimated by approximately 5$\%$. The effect of this has been investigated by scaling the PSF up by this amount and re-fitting the Crab Nebula.  The maximum effect on the flux is $\sim 5\%$, occurring at the lowest energies (see Figure \ref{fig:sysboth}). At the highest energies this effect is almost completely negligible. 





\subsubsection{Late light simulation}

This was the  largest source of uncertainty ($\sim$40\% in flux) in \cite{Abeysekara2017} and arose from a mis-modeling of the late light in the air shower. This is thought to stem from a discrepancy between the time width of the laser pulse used for calibration and the time structure of the actual shower. From simulation, it is expected that the width of the arrival time distribution of single photoelectrons (PEs) at the PMT should be $\lesssim$10 ns, but examining the raw PE distributions in data shows a discrepancy above 50 PEs.  Improved studies of the PMTs have decreased the size of this uncertainty in this work, although it is still one of the dominant sources of uncertainty.  Systematic uncertainties have been derived by varying the size of this effect and observing the impact on the flux.

\subsubsection{Charge uncertainty}
The charge uncertainty encapsulates how much a PMT measurement will vary for a fixed amount of light, and also the relative differences in photon detection efficiency from PMT to PMT. The amount of uncertainty has been varied and the effect on the flux studied. This is not a dominant source of systematic uncertainty.

\subsubsection{Absolute PMT efficiency/Time dependence}

The absolute PMT efficiency cannot be precisely determined using the calibration system (see \cite{Abeysekara2017} for a discussion). Instead, an event selection based on charge and timing cuts is implemented to identify incident vertical muons. Vertical muons provide a mono-energetic source of light and can be used to measure the relative efficiency of each PMT by matching the muon peak position to the expected from the MC simulations. These efficiencies were determined for different epochs in time and used to measure the range of uncertainties. This is one of the dominant sources of uncertainty, along with the late light simulation. 


\subsubsection{PMT threshold}

The PMT threshold (the lowest charge that a PMT can detect) is set at 0.2 PE in simulation. However, from looking at the cosmic-ray rate, the $\pm 1\sigma$ uncertainty in this may be $\pm$0.05 PE.  Simulations have been created with the PMT threshold set at 0.15 and 0.25 PEs; the effect on the flux can be seen in Figure \ref{fig:sysboth}.


\subsubsection{Bin selection}

The 2D binning scheme introduces an additional systematic uncertainty not present in \cite{Abeysekara2013}. Recall that there are 108 2D $\mathcal{B}$/energy bins, not all of which are used in the analysis. Roughly half of these bins are unpopulated.

\deleted{The bins used in the fit were chosen \textit{a priori} by looking at the distribution of estimated energies across each simulated $\mathcal{B}$ bin and keeping the central 99$\%$ of the events. This removes empty bins as well as the tails of the distribution, where statistics are low and there is more likely to be mismodeled events and a data/Monte Carlo discrepancy.}

To investigate any effect on the spectrum, the likelihood fit was repeated including the less-populated bins.  This is found to be a negligible source of systematic uncertainty.

\subsubsection{Additional sources of systematic uncertainty}\label{sec:additional}

The systematic bands for the GP and NN spectral fits shown in Figure \ref{fig:spectrum} have an additional 10$\%$ uncertainty added in quadrature with the sources of uncertainty described above. This is meant to conservatively cover a variety of systematic uncertainties stemming from detector and analysis method effects not mentioned here.  Examples include the interaction model chosen in CORSIKA and variations in the barometric pressure over time. Such changes would cause a time variation in the detector trigger rate, which would in turn have an effect on the rate of background (hadronic) events. 

\section{Discussion and Conclusions}\label{sec:disc}

\subsection{Agreement between the estimators}

The two energy estimators take very different approaches in deriving an estimate of the gamma-ray energy, and give compatible results.  There is a small discrepancy above $\sim$90 TeV, where the GP and NN forward-folded global best fits do not agree within systematic uncertainties.  However, the flux points, which show by how much the data agree with the forward-folded fit at a given energy, agree within statistical uncertainties. 

The disagreement in the forward-folded fit is likely a combination of two effects. First, potentially mis-modeled parameters in the Monte Carlo simulation may affect the two energy estimators differently. Second, due to the gamma/hadron cuts chosen, the two estimators have different efficiencies to gamma rays above 100 TeV, with the NN's being slightly higher. Given the small number of events above this energy, the inclusion or \replaced{inclusion}{exclusion} of a single event can easily account for the difference in significance.  This disagreement does not affect the significance of the observed high-energy emission from the Crab Nebula.  Regardless of which analysis technique is used, the Crab Nebula is seen \replaced{with very high significance above 56 TeV}{past 100 TeV}. 

\subsection{Comparison to other experiments}

The energy resolution is log-normal with a linear equivalent of 40$\%$(NN)-55$\%$(GP) at 1 TeV and 23$\%$(NN)-30$\%$(GP) at 50 TeV. This is a significant improvement over the previously published HAWC analysis (see Figure 2 of \cite{Abeysekara2017}) For comparison, IACTs typically have a resolution of $\sim$8-15$\%$ at 1 TeV and $\sim$15-35$\%$ at 50 TeV~\citep{Aleksic2012,Parsons2014,Park2015}. Starting around 50 TeV, the techniques presented here give comparable energy resolution to what IACTs achieve. 

There is good agreement between the spectra presented here and results from other experiments, as can be seen in Figure \ref{fig:spectrum}. This is true regardless of which energy estimator is chosen. In particular, improved detector modeling has eliminated the tension at the low-energy end ($\sim$1 TeV) between the original HAWC Crab fit presented in \cite{Abeysekara2017} and measurements from IACTs. 

Compared to the Inverse Compton model in \cite{Meyer2010}, which has been used as a reference spectrum to compare the energy scale of IACTs, both methods presented here have a 20$\%$ higher flux at 7 TeV. When applying a scaling of 0.94 on the energy scale, a deviation less than 10$\%$ from the IC model is achieved below 20 TeV and 100 TeV for the ground parameter and neural network methods respectively. The more curved spectrum of the ground parameter method tends more towards the recent publication by MAGIC~\citep{Aleksic2015}
  
\subsection{Conclusions}
This detection is the highest-energy observation of the Crab Nebula to date. Additionally, the development of two methods to identify gamma rays above 100 TeV lays the foundation for future high energy analyses across the entire HAWC field-of-view. Extending the measured energy range of previously discovered sources up to 100 TeV or higher may allow us to distinguish between leptonic and hadronic gamma ray emission mechanisms, as they have different signatures. This, in turn, may help determine if any Galactic gamma-ray sources are good candidates to be the source of the astrophysical neutrinos discovered by IceCube \citep{Aartsen2013}. Due to gamma-ray attenuation, it is expected that $>$50 TeV gamma rays will only arrive at Earth from nearby sources ($<$ 100 Mpc), excluding nearly all AGN and cosmological sources \citep{Hoffman2009}. Extending the spectra to high energies may also identify PeVatron candidates and give insight into the origins of cosmic rays \citep{Gabici2007}.

Additionally, high-energy observations also naturally lead to studies of Lorentz-invariance violation. Particle-physics models that add Lorentz-invariance-violating terms to the electromagnetic part of the Standard Model Lagrangian allow photon decay to electron/positron pairs above some energy. Since the decay probability is very nearly one for photons propagating across astrophysical distance scales, observations of high-energy photons constrain the energy at which such decay becomes allowed \citep{Martinez-Huerta2017}.
The measurements presented in this paper do not by themselves imply a limit on this energy scale; rather it must be shown that there is a statistically significant excess of events above some reconstructed energy compared to the event rate expected due to hadronic events and lower-energy photons whose energies are overestimated. Both the uncertainty on the true spectrum of the source, as well as the systematic uncertainties of the HAWC detector, must be considered. Such an analysis will be carried out in a future paper.

HAWC recently obtained a boost in high-energy sensitivity with the completion of an upgrade. This sparsely populated ``outrigger" array allows for better reconstruction of the largest, most energetic events \citep{Joshi2017}.  Data from the outrigger array is not used here but will be used in future analyses. 

\acknowledgments

We acknowledge the support from: the US National Science Foundation (NSF); the US Department of Energy Office of High-Energy Physics; 
the Laboratory Directed Research and Development (LDRD) program of Los Alamos National Laboratory; 
Consejo Nacional de Ciencia y Tecnolog\'{\i}a (CONACyT), M{\'e}xico (grants 271051, 232656, 260378, 179588, 239762, 254964, 271737, 258865, 243290, 132197, 281653)(C{\'a}tedras 873, 1563, 341), Laboratorio Nacional HAWC de rayos gamma; 
L'OREAL Fellowship for Women in Science 2014; 
Red HAWC, M{\'e}xico; 
DGAPA-UNAM (grants AG100317, IN111315, IN111716-3, IA102715, IN109916, IA102019, IN112218); 
VIEP-BUAP; 
PIFI 2012, 2013, PROFOCIE 2014, 2015; 
the University of Wisconsin Alumni Research Foundation; 
the Institute of Geophysics, Planetary Physics, and Signatures at Los Alamos National Laboratory; 
Polish Science Centre grant DEC-2014/13/B/ST9/945, DEC-2017/27/B/ST9/02272; 
Coordinaci{\'o}n de la Investigaci{\'o}n Cient\'{\i}fica de la Universidad Michoacana; Royal Society - Newton Advanced Fellowship 180385. Thanks to Scott Delay, Luciano D\'{\i}az and Eduardo Murrieta for technical support.

\bibliography{CrabPaperBib}

\end{document}